\RequirePackage{fix-cm}
\documentclass[smallextended]{svjour3}       
\smartqed  
\usepackage{graphicx,amssymb,amsfonts,mathrsfs,latexsym,amsmath,amssymb}
\usepackage{color}
\journalname{Designs, Codes and Cryptography}
\begin{document}

\title{Cyclic codes over $\mathbb{Z}_4[u]/\langle u^k\rangle$ of odd length
} \subtitle{}

\titlerunning{Cyclic codes over $\mathbb{Z}_4[u]/\langle u^k\rangle$ of odd length}

\author{Yuan Cao$^1$ $\cdot$ Qingguo Li$^1$}


\institute{Yuan Cao\at
             \email{yuan$_{-}$cao@hnu.edu.cn}\\
          Qingguo Li\at
             \email{liqingguoli@aliyun.cn}\\
           $^1$  College of Mathematics and Econometrics, Hunan University, Changsha 410082, China
          }

\date{Received: date / Accepted: date}

\maketitle

\begin{abstract}
Let $R=\mathbb{Z}_{4}[u]/\langle u^k\rangle=\mathbb{Z}_{4}
+u\mathbb{Z}_{4}+\ldots+u^{k-1}\mathbb{Z}_{4}$ ($u^k=0$) where $k\in \mathbb{Z}^{+}$ satisfies $k\geq 2$.
For any odd positive integer $n$, it is known that cyclic codes over $R$ of length $n$ are identified with ideals of the ring
$R[x]/\langle x^{n}-1\rangle$. In this paper, an explicit representation for each cyclic code over $R$ of length $n$ is provided and a formula to count the number of codewords in each code is given. Then a formula to calculate the number of cyclic codes over $R$ of length $n$
is obtained. Precisely, the dual code of each cyclic code and self-dual cyclic codes over $R$ of length $n$ are investigated.
When $k=4$, some optimal quasi-cyclic codes over $\mathbb{Z}_{4}$ of length $28$ and index $4$ are obtained from cyclic codes over $R=\mathbb{Z}_{4}
[u]/\langle u^4\rangle$.

\end{abstract}

\section{Introduction and preliminaries}
\label{intro}
Cyclic codes are one of the most important and most intensively studied classes
of linear codes with rich algebraic structure due to their representation as the
ideals of a polynomial ring. Another class of codes that became the subject
of much researcher in recent years is codes over rings. The structure of cyclic
codes over various rings is investigated by many authors (e.g. [1], [3], [4], [10]).

\par
   Let $\mathbb{Z}_{4}=\{0,1,2,3\}$ be the residue class ring of the integer ring
$\mathbb{Z}$ modulo $4$. In [13], Yildiz and Karadeniz systematically studied MacWilliams
identities, projections, and formally self-dual codes for linear codes over the
ring $\mathbb{Z}_4+u\mathbb{Z}_4$ ($u^2=0$) and their application to real and complex lattices. In [14], Yildiz and Aydin investigated algebraic structure of
cyclic codes over the ring $\mathbb{Z}_4+u\mathbb{Z}_4$.
Making use of the structure of these cyclic codes, Yildiz
and Aydin conducted a computer search and obtained some new linear codes
over $\mathbb{Z}_4$. In [8], cyclic codes over $\mathbb{Z}_q + u\mathbb{Z}_q$ ($u^2=0$) are investigated
where $q$ is power of a prime.

\par
  In this paper, let $k\geq 2$ be an arbitrary integer and denote
\begin{eqnarray*}
R&=&\mathbb{Z}_{4}[u]/\langle u^k\rangle=\{\sum_{i=0}^{k-1}a_iu^i\mid a_0,a_1,\ldots,a_{k-1}\in \mathbb{Z}_{4}\}\\
 &=&\mathbb{Z}_{4}+u\mathbb{Z}_{4}+\ldots+u^{k-1}\mathbb{Z}_{4} \ (u^k=0)
\end{eqnarray*}

\noindent
  It is well-known that $R$ is a finite commutative local ring with the maximal ideal
$\langle 2,u\rangle=2R+uR$ and $R/\langle 2,u\rangle\cong \mathbb{F}_2$ which is a finite field of two elements.

\par
  For any positive integer $n$,
let $R[x]/\langle x^n-1\rangle$ be the residue class ring of the polynomial ring $R[x]$ modulo its ideal $\langle x^n-1\rangle$ generated by $x^n-1$.
Then every element of $R[x]/\langle x^n-1\rangle$ can be uniquely expressed as
$a(x)+\langle x^n-1\rangle$ where $a(x)=\sum_{j=0}^{n-1}a_jx^j\in R[x]$ with $a_0,a_1,\ldots,a_{n-1}\in R$.
As usual, we will identify $a(x)+\langle x^N-1\rangle$ with $a(x)$ for simplicity. Hence
\begin{center}
$R[x]/\langle x^n-1\rangle=\{\sum_{j=0}^{n-1}a_jx^j\mid a_0,a_1,\ldots,a_{n-1}\in R\}$
\end{center}
which is a ring with the operations
defined by the usual polynomial operations modulo $x^n-1$.
It is well-known that cyclic codes over
$R$ of length $n$ are identified with ideals of the ring $R[x]/\langle x^n-1\rangle$, under the
identification map $\theta: R^n \rightarrow R[x]/\langle x^n-1\rangle$ via $\theta: (a_0,a_1,\ldots,a_{n-1})\mapsto
\sum_{j=0}^{n-1}a_jx^j$ ($\forall a_0,a_1,\ldots,a_{n-1}\in R$).

\par
  When $k\geq 3$ arbitrary, the following questions for cyclic codes over $R$ have not been
solved completely to the best of our knowledge:

\vskip 2mm\par
  $\diamondsuit$ For each cyclic code ${\cal C}$ over $R$ of length $N$, give a unique representation for
${\cal C}$ and provide a clear formula to count the number of codewords in ${\cal C}$. Then obtain a formula to count the number of all such cyclic codes.

\vskip 2mm\par
  $\diamondsuit$ Using the representation of any cyclic code ${\cal C}$, give the representation
of the dual code of ${\cal C}$ precisely and determine the self-duality of ${\cal C}$.

\vskip 2mm\par
   From now on, let $k\geq 2$ and $n$ be an odd positive integer. We focus our attention to consider cyclic codes over $R$ of length $n$ in this paper.

\vskip 3mm\par
   The present paper is organized as follows. In Section 2, we sketch the basic theory of Galois rings of characteristic $4$.
In Section 3, we provide an explicit representation for each cyclic code over $R$ of length $n$ and give a formula to count the number of codewords in each code. As a corollary, we obtain a formula for the number of all such cyclic codes. In Section 4, we determine the dual code of each cyclic code over $R$ of length $n$ and investigate self-duality for such codes. In Section 5, we provide a generator matrix for quasi-cyclic codes over $\mathbb{Z}_4$ of length $28$
obtained from cyclic codes over $\mathbb{Z}_4[u]/\langle u^4\rangle$.
Finally, we list all $293687$ cyclic codes and all $791$ self-dual cyclic codes of length $7$ over $\mathbb{Z}_4[u]/\langle u^4\rangle$, and give optimal quasi-cyclic codes over $\mathbb{Z}_4$ of length $28$ and index $4$ from
these cyclic codes over $\mathbb{Z}_4[u]/\langle u^4\rangle$ in Section 6.

\section{Preliminaries}
In this section, we sketch the basic theory of Galois rings of characteristic $4$, which are
Galois extension rings of $\mathbb{Z}_4$.
Readers are referred to Wan [11] and [12]
for more details on Galois rings.

\par
   Let $\mathbb{Z}_4[x]$ be the polynomial ring in an indeterminate $x$ over $\mathbb{Z}_4$ and
$\mathbb{F}_2=\{0,1\}$ bing the field of two elements. It is well known that
the map $^{-}: \mathbb{Z}_4\rightarrow \mathbb{F}_2$ defined by $\overline{0}=\overline{2}=0$ and $\overline{1}=\overline{3}=1$
is a surjective ring homomorphism and can be naturally extended to a surjective ring isomorphism from
$\mathbb{Z}_4[x]$ onto $\mathbb{F}_2[x]$ as follows:
\begin{equation}
^{-}: a(x)\mapsto \overline{a}(x)=\sum_{i=0}^s\overline{a}_ix^i \ (\forall a(x)=\sum_{i=0}^sa_ix^i\in \mathbb{Z}_4[x]
\ {\rm where} \ a_i\in \mathbb{Z}_4).
\end{equation}

\par
   Recall that a \textit{basic irreducible} (\textit{basic primitive}) monic polynomial $f(x)$ of degree $d$ in $\mathbb{Z}_4[x]$ is a monic
polynomial of degree $d$ over $\mathbb{Z}_4$ such that $\overline{f}(x)$ is irreducible (primitive) over $\mathbb{F}_2$.
In the following, let $f(x)$ be a fixed basic irreducible monic polynomial of degree $d$ in $\mathbb{Z}_4[x]$.
We consider the residue class ring
$$\mathcal{K}=\mathbb{Z}_4[x]/\langle f(x)\rangle$$
of $\mathbb{Z}_4[x]$ modulo its ideal $\langle f(x)\rangle$ generated by $f(x)$. $\mathcal{K}$ is a \textit{Galois ring}
of characteristic $4$ and $4^d$ elements (cf. [11] Theorem 6.1).
  It is clear that each element $\alpha$ of $\mathcal{K}$ can be uniquely expressed as
$\alpha=\sum_{i=0}^{d-1}a_ix^i$, $a_0,a_1,\ldots,a_{d-1}\in \mathbb{Z}_4.$
  In this paper, we set
\begin{equation}
\mathcal{K}=\{\sum_{i=0}^{d-1}a_ix^i\mid a_0,a_1,\ldots,a_{d-1}\in \mathbb{Z}_4\}.
\end{equation}
For any $\beta=\sum_{i=0}^{d-1}b_ix^i$ where $b_0,b_1,\ldots,b_{d-1}\in \mathbb{Z}_4$,
the addition $\alpha+\beta$ and multiplication $\alpha\beta$ of $\alpha$ and $\beta$ in the ring $\mathcal{K}$ are given as follows:
$$\alpha+\beta=\sum_{i=0}^{d-1}(a_i+b_i)x^i,
\ \alpha\beta\equiv (\sum_{i=0}^{d-1}a_ix^i)(\sum_{i=0}^{d-1}b_ix^i) \ ({\rm mod} \ f(x)),$$
where $(\sum_{i=0}^{d-1}a_ix^i)(\sum_{i=0}^{d-1}b_ix^i)$ is the usual product of
polynomials $\sum_{i=0}^{d-1}a_ix^i$ and $\sum_{i=0}^{d-1}b_ix^i$
in $\mathbb{Z}_4[x]$.

\par
  Similar, in this paper we set
\begin{equation}
\mathcal{F}=\mathbb{F}_2[x]/\langle \overline{f}(x)\rangle=\{\sum_{i=0}^{d-1}g_ix^i\mid g_0,g_1,\ldots,g_{d-1}\in \mathbb{F}_2\},
\end{equation}
which is a finite field of $2^d$ elements since $\overline{f}(x)$ is an irreducible polynomial of
degree $d$ in $\mathbb{F}_2[x]$.
For any $\xi=\sum_{i=0}^{d-1}g_ix^i$ and $\eta=\sum_{i=0}^{d-1}h_ix^i$ where $g_i,h_i\in \mathbb{F}_2$ for all $i=0,1,\ldots,d-1$,
the addition $\xi+\eta$ and multiplication $\xi\eta$ of $\xi$ and $\eta$ in the field $\mathcal{F}$ are given as follows:
$$\xi+\eta=\sum_{i=0}^{d-1}(g_i+h_i)x^i,
\ \xi\eta\equiv (\sum_{i=0}^{d-1}g_ix^i)(\sum_{i=0}^{d-1}h_ix^i) \ ({\rm mod} \ \overline{f}(x)),$$
where $(\sum_{i=0}^{d-1}g_ix^i)(\sum_{i=0}^{d-1}h_ix^i)$ is the usual product of polynomials
in $\mathbb{F}_2[x]$.

\par
   Since the kernel of the surjective ring homomorphism $^{-}$ defined by Equation (1) is the
ideal $2(\mathbb{Z}_4[x])$ and the image of $f(x)(\mathbb{Z}_4[x])$ under $^{-}$ is $\overline{f}(x)(\mathbb{F}_2[x])$,
the ring homomorphism $^{-}$ induces a surjective ring homomorphism from $\mathcal{K}$ onto $\mathcal{F}$, which is still denoted by $^{-}$
in this paper.
Precisely, we have
$$^{-}: \mathcal{K}\rightarrow \mathcal{F} \ {\rm via} \  a(x)\mapsto \overline{a}(x)=\sum_{i=0}^{d-1}\overline{a}_ix^i$$
for any $a(x)=\sum_{i=0}^{d-1}a_ix^i\in \mathcal{K}$ with $a_0,a_1,\ldots,a_{d-1}\in \mathbb{Z}_4$.

\par
   As $\mathbb{F}_2=\{0,1\}$ and $\mathbb{Z}_4=\{0,1,2,3\}$, we can regard $\mathbb{F}_2$ as a subset of $\mathbb{Z}_4$.
But it is well known that $\mathbb{F}_2$ is not a subring of $\mathbb{Z}_4$. For example, $1+1=0$ in $\mathbb{F}_2$ but
$1+1=2\neq 0$ in $\mathbb{Z}_4$. In this paper, we will regard $\mathbb{F}_2$ as a subset of $\mathbb{Z}_4$ for convenience.
In this sense, each element of $\mathbb{Z}_4$ has the following $2$-adic expansion uniquely:
\begin{equation}
t_0+2t_1, \ {\rm where} \ t_0,t_1\in \mathbb{F}_2.
\end{equation}
It is clear that $\overline{t_0+2t_1}=t_0$ for any $t_0,t_1\in \mathbb{F}_2$.

\par
   Similarly, using Equations (2) and (3) we will regard $\mathcal{F}$ as a subset of $\mathcal{K}$ in this paper.
However, readers need to notice that $\mathcal{F}$ is not a subring of $\mathcal{K}$.
In this sense, by Equation (4)
we conclude that each element of $\mathcal{K}$ has the following $2$-adic expansion uniquely:
$$\xi_0+2\xi_1, \ {\rm where} \ \xi_0,\xi_1\in \mathcal{F}.$$
It is clear that $\overline{\xi_0+2\xi_1}=\xi_0$ for any $\xi_0,\xi_1\in \mathcal{F}$.

\section{Representation for all distinct cyclic codes over $R$ of length $n$}
In this section, we investigate the structures of $R[x]/\langle x^{n}-1\rangle$ first. Then
give an explicit presentation for all distinct cyclic codes over $R$ of length $n$. In the reset of this paper, we denote

\par
  $\bullet$ ${\cal A}=\mathbb{Z}_{4}[x]/\langle x^{n}-1\rangle=\{\sum_{i=0}^{n-1}a_ix^i\mid a_i\in \mathbb{Z}_4\}$,

\noindent
which is a finite commutative ring with identity under the usual polynomial polynomial addition and
multiplication modulo $x^n-1$.

\par
  As ${\cal A}$ is a commutative ring with identity, we can form the polynomial ring
$${\cal A}[u]=\{\sum_{j=0}^s\xi_ju^j\mid \xi_j\in {\cal A}, \ j=0,1,\ldots,s,\ s\geq 0\},$$
where $u$ is an indeterminate on ${\cal A}$. Let $\langle u^k\rangle={\cal A}u^k$, which is the ideal
of ${\cal A}[u]$ generated by $u^k$. Then we have the following residue class ring of ${\cal A}[u]$ modulo
its ideal $\langle u^k\rangle$:

\vskip 2mm \par
  $\bullet$ ${\cal A}[u]/\langle u^k\rangle={\cal A}+u{\cal A}+\ldots+u^{k-1}{\cal A}=\{\sum_{j=0}^{k-1}\xi_ju^j\mid \xi_j\in {\cal A}, \ j=0,1,\ldots,k-1\}$ ($u^k=0$).

\vskip 3mm
\par
   First, we build a relationship between rings ${\cal A}[u]/\langle u^k\rangle$ and
$R[x]/\langle x^n-1\rangle$. Let $\xi=\sum_{j=0}^{k-1}\xi_ju^j\in {\cal A}[u]/\langle u^k\rangle$ with
$\xi_j\in {\cal A}$. As ${\cal A}=\mathbb{Z}_4[x]/\langle x^n-1\rangle$, each $\xi_j$ is uniquely expressed as
$$\xi_j=a_{0,j}+a_{1,j}x+\ldots+a_{n-1,j}x^{n-1}$$
where $a_{i,j}\in \mathbb{Z}_4$ for all $i=0,1,\ldots,n-1$ and $j=0,1,\ldots,k-1$. Hence
$$\xi=(1,x,\ldots,x^{n-1})\left(\begin{array}{cccc}a_{0,0} & a_{0,1} & \ldots & a_{0,k-1}
\cr a_{1,0} & a_{1,1} & \ldots & a_{1,k-1} \cr \ldots & \ldots & \ldots & \ldots \cr a_{n-1,0} & a_{n-1,1} & \ldots & a_{n-1,k-1}\end{array}\right)\left(\begin{array}{c}1\cr u\cr \ldots \cr u^{k-1}\end{array}\right).$$
Define a natural map

\par
  $\bullet$ $\Psi: {\cal A}[u]/\langle u^k\rangle\rightarrow R[x]/\langle x^n-1\rangle$ by
$$\Psi(\xi)=\sum_{i=0}^{n-1}\alpha_ix^i, \ {\rm where} \ \alpha_i=\sum_{j=0}^{k-1}a_{i,j}u^k\in R, \ i=0,1,\ldots, n-1.$$
Then by a direct
calculation, one can easily verify the following lemma.

\vskip 3mm
\noindent
  {\bf Lemma 3.1} \textit{The map $\Psi$ defined above is a ring isomorphism from
${\cal A}[u]/\langle u^k\rangle$ onto $R[x]/\langle x^{n}-1\rangle$}.

\vskip 3mm
\par
   By Lemma 3.1, $\mathcal{C}$ is a cyclic code over $R$ of length $n$, i.e., $\mathcal{C}$ is an ideal of the ring
$R[x]/\langle x^{n}-1\rangle$ if and only if the is a unique ideal $J$ of the ring ${\cal A}[u]/\langle u^k\rangle$
such that $\Psi(J)=\mathcal{C}$. Therefore, in order to determine all cyclic codes over $R$ of length $n$, it is sufficient to determine all ideals of ${\cal A}[u]/\langle u^k\rangle$. To do this, we need to investigate the structure and properties of the ring ${\cal A}$.

\par
   As $n$ is odd, by [11] Proposition 7.17(i) there are pairwise coprime monic basic irreducible polynomials $f_1(x),\ldots,f_r(x)$ in $\mathbb{Z}_{4}[x]$
such that
\begin{center}
$x^n-1=f_1(x)\ldots f_r(x)$.
\end{center}
Let $1\leq j\leq r$. We assume ${\rm deg}(f_j(x))=d_j$ and denote $F_j(x)=\frac{x^{n}-1}{f_j(x)}\in \mathbb{Z}_{4}[x]$. Then $F_j(x)$ and $f_j(x)$
are coprime in $\mathbb{Z}_{4}[x]$. By [11] Proposition 7.17(ii)
there exist $v_j(x),w_j(x)\in\mathbb{Z}_{4}[x]$ such that
\begin{equation}
v_j(x)F_j(x)+w_j(x)f_j(x)=1.
\end{equation}
From now on, we denote

\par
  $\bullet$ ${\cal K}_j=\mathbb{Z}_{4}[x]/\langle f_j(x)\rangle=\{\sum_{i=0}^{d_i-1}a_ix^i\mid a_0,a_1,\ldots,a_{d_j-1}\in
\mathbb{Z}_4\}$, which is a Galois ring of characteristic $4$ and
cardinality $4^{d_j}$ by Section 2.

\par
  $\bullet$ $e_j(x)\in {\cal A}$ defined by $e_j(x)\equiv v_j(x)F_j(x)=1-w_j(x)f_j(x)$ $({\rm mod} \ x^{n}-1)$.

\vskip 2mm
   Then by [11] Chapter 7 we have the following conclusions.

\vskip 3mm
\noindent
  {\bf Lemma 3.2} \textit{Using the notations above, we have the following}:

\vskip 2mm\par
  (i) (cf. [11] Theorem 7.14 (i) and (ii)) \textit{$e_1(x)+\ldots+e_r(x)=1$, $e_j(x)^2=e_j(x)$
and $e_j(x)e_l(x)=0$  in the ring ${\cal A}$ for all $1\leq j\neq l\leq r$}.

\vskip 2mm\par
  (ii) (cf. [11] Theorem 7.14 (iii) and (iv)) \textit{${\cal A}={\cal A}_1\oplus\ldots \oplus{\cal A}_r$ where ${\cal A}_j={\cal A}e_j(x)$ with
$e_j(x)$ as its multiplicative identity and satisfies ${\cal A}_j{\cal A}_l=\{0\}$ for all $1\leq j\neq l\leq r$}.

\vskip 2mm\par
  (iii) (cf. [11] Corollary 7.15) \textit{For any integer $j$, $1\leq j\leq r$, define $\varphi_j: a(x)\mapsto e_j(x)a(x)$ $(${\rm mod} $x^{n}-1)$
for any $a(x)\in {\cal K}_j$. Then $\varphi_j$ is a ring isomorphism from ${\cal K}_j$ onto ${\cal A}_j$. Furthermore, the follow map}
$$\varphi(a_1(x),\ldots,a_r(x))=\sum_{j=1}^r\varphi_j(a_j(x))=\sum_{j=1}^re_j(x)a_j(x) \ ({\rm mod} \ x^n-1)$$
\textit{$(\forall a_j(x)\in {\cal K}_j, \ j=1,\ldots,r)$ is a ring isomorphism from
the direct product ring ${\cal K}_1\times \ldots \times {\cal K}_r$ onto $\mathcal{A}$}.

\vskip 2mm\par
  (iv) (cf. [11] Section 7.2) \textit{For any integer $j$, $1\leq j\leq r$, ${\cal A}_j={\cal A}\frac{x^n-1}{f_j(x)}$ which is a cyclic code over $\mathbb{Z}_{4}$
of length $n$ with parity check polynomial $f_j(x)$}.

\vskip 3mm\par
   For each $1\leq j\leq r$, let ${\cal K}_j[u]$ be the polynomial ring
over the Galois ring with the indeterminate $u$, and denote the residue class ring
of ${\cal K}_j[u]$ modulo its ideal generated by $u^k$ as the following

\vskip 2mm\par
  $\bullet$ ${\cal K}_j[u]/\langle u^k\rangle={\cal K}_j+u{\cal K}_j+\ldots+u^{k-1}{\cal K}_j$ ($u^k=0$).

\vskip 2mm\noindent
 Then we have the following conclusion for the relationship between ideals of rings ${\cal K}_j[u]/\langle u^k\rangle$,
$1\leq j\leq r$,
and ideals of the ring ${\cal A}[u]/\langle u^k\rangle$.

\vskip 3mm
\noindent
  {\bf Lemma 3.3} \textit{Let $\emptyset\neq C\subseteq {\cal A}[u]/\langle u^k\rangle$. Then $C$ is an ideal
of ${\cal A}[u]/\langle u^k\rangle$ if and only if for each
integer $j$, $1\leq j\leq r$, there is a unique ideal $C_j$ of the ring ${\cal K}_j[u]/\langle u^k\rangle$
such that}
$$C=\bigoplus_{j=1}^re_j(x)C_j,$$
\textit{where $e_j(x)C_j=\{e_j(x)\beta_j  \ ({\rm mod} \ x^n-1)\mid \beta_j\in C_j\}\subseteq \mathcal{A}[u]/\langle u^k\rangle$ for all $j=1,\ldots,r$.
Moreover, the number of elements in $C$ is equal to $|C|=\prod_{j=1}^r|C_j|$}.

\vskip 3mm
\noindent
  \textit{Proof} By Lemma 3.2(iii), the ring isomorphism $\varphi$ from ${\cal K}_1\times\ldots\times {\cal K}_r$
onto ${\cal A}$ induces a polynomial ring isomorphism $\Phi_0$ from ${\cal K}_1[u]\times\ldots\times {\cal K}_r[u]$
onto ${\cal A}[u]$ in the natural way:
\begin{eqnarray*}
&&\Phi_0\left(\sum_{i}u^ia_{1,i}(x),\ldots,\sum_{i}u^ia_{r,i}(x)\right)=\sum_{i}u^i\varphi(a_{1,i}(x),\ldots,a_{r,i}(x))\\
&=&\sum_{i}u^i(e_1(x)a_{1,i}(x)+\ldots+e_r(x)a_{r,i}(x) \ ({\rm mod} \ x^n-1))
\end{eqnarray*}
for all $\sum_{i}u^ia_{1,i}(x)\in {\cal K}_1[u], \ldots, \sum_{i}u^ia_{r,i}(x)\in {\cal K}_r[u]$. Therefore, $\Phi_0$
induces a ring isomorphism $\Phi$ from
$({\cal K}_1[u]/\langle u^k\rangle)\times\ldots\times {\cal K}_r[u]/\langle u^k\rangle$
onto ${\cal A}[u]/\langle u^k\rangle$ in the natural way:
\begin{eqnarray*}
\Phi(\beta_1,\ldots,\beta_r)&=&\sum_{l=0}^{k-1}\varphi(a_{l,1}(x),\ldots,a_{l,r}(x))u^l\\
  &=&\sum_{l=0}^{k-1}u^l\left(\sum_{j=1}^re_j(x)a_{l,j}(x) \ ({\rm mod} \ x^n-1)\right)\\
  &=&e_1(x)\beta_1+\ldots+e_r(x)\beta_r(x) \ ({\rm mod} \ x^n-1).
\end{eqnarray*}
for any $\beta_j=\sum_{l=0}^{k-1}a_{l,j}(x)u^l\in {\cal K}_j[u]/\langle u^k\rangle$
where $a_{l,j}(x)\in {\cal K}_j$ for all $l=0,1,\ldots,k-1$ and $j=1,\ldots,r$.

\par
   By properties for the direct product of rings and isomorphic rings, we see that $C$ is an ideal of ${\cal A}[u]/\langle u^k\rangle$
if and only if for each
integer $j$, $1\leq j\leq r$, there is a unique ideal $C_j$ of the ring ${\cal K}_j[u]/\langle u^k\rangle$ such that
\begin{eqnarray*}
C&=&\Phi(C_1\times\ldots\times C_r)=\Phi\{(\beta_1,\ldots,\beta_r)\mid \beta_j\in C_j, \ j=1,\ldots,r\}\\
  &=&\{\Phi(\beta_1,\ldots,\beta_r)\mid \beta_j\in C_j, \ j=1,\ldots,r\}\\
  &=&\{\sum_{j=1}^re_j(x)\beta_j \ ({\rm mod} \ x^n-1)\mid \beta_j\in C_j, \ j=1,\ldots,r\}\\
  &=&\bigoplus_{j=1}^re_j(x)C_j \ ({\rm mod} \ x^n-1).
\end{eqnarray*}
Moreover, the elements contained in $C$ is equal to
$|C|=|C_1\times\ldots\times C_r|=\prod_{j=0}^r|C_j|$.
\hfill $\Box$

\vskip 3mm
\par
   Next, we give a direct sum decomposition for any
cyclic code over $R$ of length $n$.

\vskip 3mm
\noindent
  {\bf Theorem 3.4} \textit{Let $\emptyset\neq \mathcal{C}\subseteq R[x]/\langle x^n-1\rangle$.
Then $\mathcal{C}$ is a cyclic code over $R$ of length $n$ if and only if for each
integer $j$, $1\leq j\leq r$, there is a unique ideal $C_j$ of the ring ${\cal K}_j[u]/\langle u^k\rangle$
such that}
$$\mathcal{C}=\bigoplus_{j=1}^r\Psi(e_j(x)C_j).$$
\textit{Moreover, the number of codewords in $\mathcal{C}$ is equal to $|\mathcal{C}|=\prod_{j=0}^r|C_j|$}.

\vskip 3mm
\noindent
  \textit{Proof} By Lemmas 3.1, $\Psi$ is a ring isomorphism from
${\cal A}[u]/\langle u^k\rangle$ onto $R[x]/\langle x^n-1\rangle$.
Hence every cyclic code $\mathcal{C}$ over $R$ of length $n$, i.e.,
$\mathcal{C}$ is an ideal of $R[x]/\langle x^n-1\rangle$, can be uniquely
expressed as
\begin{center}
$\mathcal{C}=\Psi(C)$, where $C$ is an ideal of ${\cal A}[u]/\langle u^k\rangle$.
\end{center}
Then the conclusion follows from Lemma 3.3 immediately.
\hfill $\Box$

\vskip 3mm
\noindent
  {\bf Remark} Using the notations of Theorem 3.4, ${\cal C}=\oplus_{j=1}^r\Psi(e_j(x)C_j)$, where $C_j$
is an ideal of ${\cal K}_j[u]/\langle u^k\rangle$ for $j=1,\ldots,r$, is called a
\textit{canonical form decomposition} of the cyclic code ${\cal C}$ over $R$ of length $n$.

\vskip 3mm\par
  Now, we consider how to determine the ideals of ${\cal K}_j[u]/\langle u^k\rangle$ for all $j=1,\ldots,r$.
Let $1\leq j\leq r$. In the rest of this paper, we adopt the following notations:

\vskip 2mm\par
  $\bullet$ ${\cal F}_j=\mathbb{F}_{2}[x]/\langle \overline{f}_j(x)\rangle=\{\sum_{l=0}^{d_j-1}g_lx^l
\mid g_0,g_1,\ldots,g_{d_j-1}\in \mathbb{F}_{2}\}$, which is a finite field of $2^{d_j}$
elements since $\overline{f}_j(x)$ is an irreducible polynomial over $\mathbb{F}_{2}$ of degree $d_j$.

\vskip 2mm\par
  $\bullet$ ${\cal F}_j[u]/\langle u^k\rangle=\{\sum_{i=0}^{k-1}b_i(x)u^i\mid b_0(x),b_1(x),\ldots,b_{k-1}(x)\in {\cal F}_j\}
={\cal F}_j+u{\cal F}_j+\ldots+u^{k-1}{\cal F}_j$ ($u^k=0$).

\vskip 2mm\noindent
  The following lemma follows from polynomial theory over finite fields and finite chain ring theory (cf. Norton et al. [9])
immediately. Here, we omit its proof.

\vskip 3mm\noindent
  {\bf Lemma 3.5} \textit{Using the notations above, for any $1\leq j\leq r$ and $1\leq s\leq k$ we have the following}:

\par
  (i) \textit{${\cal F}_j[u]/\langle u^s\rangle={\cal F}_j+u{\cal F}_j+\ldots+u^{s-1}{\cal F}_j$ $(u^s=0)$
is a finite commutative chain ring with the unique maximal ideal $u({\cal F}_j[u]/\langle u^s\rangle)$,
the nilpotency index of $u$ is equal to $s$ and the residue class field of ${\cal F}_j[u]/\langle u^s\rangle$ is $({\cal F}_j[u]/\langle u^s\rangle)/u({\cal F}_j[u]/\langle u^s\rangle)$ which is isomorphic to ${\cal F}_j$}.

\par
  (ii) \textit{Every element $\alpha$ of ${\cal F}_j[u]/\langle u^s\rangle$
has a unique $u$-adic expansion}:
\begin{center}
$\alpha=t_0(x)+ut_1(x)+\ldots+u^{s-1}t_{s-1}(x)$, $t_0(x),t_1(x),\ldots,t_{s-1}(x)\in {\cal F}_j$.
\end{center}

\par
  (iii) \textit{All distinct ideals of ${\cal F}_j[u]/\langle u^s\rangle$ are given by:
$$u^l({\cal F}_j[u]/\langle u^s\rangle)=\{\sum_{i=l}^{s-1}b_i(x)u^i\mid b_l(x),\ldots,b_{s-1}(x)\in {\cal F}_j\},
\ 0\leq l\leq s.$$
Moreover, we have $|u^l({\cal F}_j[u]/\langle u^s\rangle)|=|{\cal F}_j|^{s-l}=2^{d_j(s-l)}$ for all $l$}.

\par
  (iv) \textit{The group of invertible elements of ${\cal F}_j[u]/\langle u^s\rangle$ is given by
\begin{center}
$({\cal F}_j[u]/\langle u^s\rangle)^{\times}=\{\sum_{l=0}^{s-1}b_l(x)u^l\mid b_0(x)\neq 0, b_0(x),b_1(x),\ldots,b_{s-1}(x)\in {\cal F}_j\}$
\end{center}
and $|({\cal F}_j[u]/\langle u^s\rangle)^{\times}|=(2^{d_j}-1)2^{(s-1)d_j}$}.

\par
  (v) \textit{For any
nonzero element $\alpha$ of ${\cal F}_j[u]/\langle u^s\rangle$ there is a unique
integer $i$, $0\leq i\leq s-1$, such that $\alpha=u^i\xi$ for some $\xi\in ({\cal F}_j[u]/\langle u^s\rangle)^{\times}$}.

\vskip 3mm\par
  As stated in Section 2,
we will regard ${\cal F}_j$ as a subset
of the Galois ring ${\cal K}_j$ (but ${\cal F}_j$ is not a subring of ${\cal K}_j$) for convenience. Then each
element of ring ${\cal K}_j$ has the following $2$-adic expansion uniquely:
$$\xi_0+2\xi_1, \ {\rm where} \ \xi_0,\xi_1\in \mathcal{F}_j.$$
Moreover, we have $\overline{\xi_0+2\xi_1}=\xi_0$ for any $\xi_0,\xi_1\in \mathcal{F}_j$.

\par
   Similarly, we will regard ${\cal F}_j[u]/\langle u^k\rangle$ as a subset
of ${\cal K}_j[u]/\langle u^k\rangle$ in this paper (but ${\cal F}_j[u]/\langle u^k\rangle$ is not a subring
of ${\cal K}_j[u]/\langle u^k\rangle$). Then one can easily deduce that each element
of ${\cal K}_j[u]/\langle u^k\rangle$ has the following $2$-adic expansion uniquely:
\begin{equation}
\eta_0+2\eta_1, \ {\rm where} \ \eta_0, \eta_1\in {\cal F}_j[u]/\langle u^k\rangle.
\end{equation}
Moreover, we define a natural map

\vskip 2mm\par
  $\bullet$  $\tau: {\cal K}_j[u]/\langle u^k\rangle\rightarrow {\cal F}_j[u]/\langle u^k\rangle
\ {\rm via} \ \tau(\eta_0+2\eta_1)=\eta_0 \ (\forall \eta_0, \eta_1\in {\cal F}_j[u]/\langle u^k\rangle)$

\vskip 2mm\noindent
Then $\tau$ is a surjective ring homomorphism from
${\cal K}_j[u]/\langle u^k\rangle$ onto ${\cal F}_j[u]/\langle u^k\rangle$ with
kernel
${\rm Ker}(\tau)=2({\cal K}_j[u]/\langle u^k\rangle)$.

\par
    Let $\xi\in {\cal F}_j[u]/\langle u^k\rangle$ and $2\eta\in 2({\cal K}_j[u]/\langle u^k\rangle)$
where $\eta\in {\cal F}_j[u]/\langle u^k\rangle$. We calculate the product $\xi\eta$ of
$\xi$ and $\eta$ in the ring ${\cal F}_j[u]/\langle u^k\rangle$. Also, we can regard $\xi\eta$ as an
element of ${\cal K}_j[u]/\langle u^k\rangle$. Then by Equation (6), we obtain a unique element $2(\xi\eta)$ in $2({\cal K}_j[u]/\langle u^k\rangle)$. Now, we define
\begin{equation}
\xi\cdot 2\eta=2(\xi\eta)\in 2({\cal K}_j[u]/\langle u^k\rangle).
\end{equation}
Especially, we have $\xi\cdot 2=2\xi\in 2({\cal K}_j[u]/\langle u^k\rangle)$ if $\eta=1$.
Then $2({\cal K}_j[u]/\langle u^k\rangle)$ is an ${\cal F}_j[u]/\langle u^k\rangle$-module. Moreover, we have

\vskip 3mm
\noindent
   {\bf Lemma 3.6} \textit{With the scalar multiplication defined by Equation $(7)$, the map
$\psi_j$ defined by
$$\psi_j(\xi)=2\xi\in {\cal K}_j[u]/\langle u^k\rangle \ (\forall \xi \in{\cal F}_j[u]/\langle u^k\rangle)$$
is an
${\cal F}_j[u]/\langle u^k\rangle$-module isomorphism from ${\cal F}_j[u]/\langle u^k\rangle$ onto
$2({\cal K}_j[u]/\langle u^k\rangle)$}.

\vskip 3mm
\noindent
   \textit{Proof} By Equation (6), every element $\vartheta$ of ${\cal K}_j[u]/\langle u^k\rangle$ is uniquely
expressed as $\vartheta=\eta_0+2\eta_1$ where $\eta_0,\eta_1\in {\cal F}_j[u]/\langle u^k\rangle$. Here we regard ${\cal F}_j[u]/\langle u^k\rangle$ as a subset of ${\cal K}_j[u]/\langle u^k\rangle$.
Then by $4=0$ in ${\cal K}_j[u]/\langle u^k\rangle$, it follows that
$2\vartheta=2\eta_0$, which implies
$\psi_j(\eta_0)=2\vartheta$. Hence the map $\psi_j$ is a bijection.

\par
  For any $\xi_1,\xi_2\in {\cal F}_j[u]/\langle u^k\rangle$,
let $\xi_1+\xi_2=\alpha$ and $\xi_1\xi_2=\beta$ as elements of the ring ${\cal F}_j[u]/\langle u^k\rangle$. Now, we regard both
$\xi_1$ and $\xi_2$ as elements of ${\cal K}_j[u]/\langle u^k\rangle$ and assume
$\xi_1+\xi_2=\zeta$ and $\xi_1\xi_2=\eta$ as elements of the ring ${\cal K}_j[u]/\langle u^k\rangle$.
Since $\tau$ is a surjective ring homomorphism from ${\cal K}_j[u]/\langle u^k\rangle$ onto ${\cal F}_j[u]/\langle u^k\rangle$,
by Equation (6) and the definition of $\tau$ we have $\tau(\zeta)=\alpha$ and $\tau(\eta)=\beta$, which
imply $\zeta=\alpha+2\alpha_1$ and $\eta=\beta+2\beta_1$ for some $\alpha_1,\beta_1\in {\cal F}_j[u]/\langle u^k\rangle$. By $4=0$
in ${\cal K}_j[u]/\langle u^k\rangle$,
we have
\begin{eqnarray*}
\psi_j(\xi_1+\xi_2)&=&\psi_j(\alpha)=2\alpha=2(\alpha+2\alpha_1)=2\zeta=2(\xi_1+\xi_2)\\
   &=&2\xi_1+2\xi_2=\psi_j(\xi_1)+\psi_j(\xi_2)
\end{eqnarray*}
 and $\psi_j(\xi_1\xi_2)=2\beta=2(\beta+2\beta_1)=2\eta=2(\xi_1\xi_2)
=\xi_1(2\xi_2)=\xi_1\psi_j(\xi_2)$. Hence $\psi_j$ is an ${\cal F}_j[u]/\langle u^k\rangle$-module isomorphism from ${\cal F}_j[u]/\langle u^k\rangle$ onto
$2({\cal K}_j[u]/\langle u^k\rangle)$.
\hfill $\Box$

\vskip 3mm
\noindent
   {\bf Remark} By Lemma 3.6, every element of $2({\cal K}_j[u]/\langle u^k\rangle)$ can be
uniquely expressed as: $2\xi$, $\xi\in {\cal F}_j[u]/\langle u^k\rangle$, where
we regard $\xi$ in the expression $2\xi$ as an element of ${\cal K}_j[u]/\langle u^k\rangle$
in the rest of this paper.

\vskip 3mm\par
  Then we present all distinct ideals
of the ring ${\cal K}_j[u]/\langle u^k\rangle$ $(1\leq j\leq r)$.

\vskip 3mm
\noindent
  {\bf Theorem 3.7} \textit{Using the notations above, all distinct ideals of ${\cal K}_j[u]/\langle u^k\rangle$ are
given by the following table}:
{\small\begin{center}
\begin{tabular}{llll}\hline
case &  number of ideals  &  $C_j$ (ideal of ${\cal K}_j[u]/\langle u^k\rangle$)    &   $|C_j|$ \\ \hline
I.   & $k+1$  & $\bullet$ $\langle u^i\rangle$ \ $(0\leq i\leq k)$ & $2^{2d_j(k-i)}$ \\
II.   & $k$     & $\bullet$  $\langle 2u^s\rangle$ \ $(0\leq s\leq k-1)$ &  $2^{d_j(k-s)}$  \\
III.   & $\Omega_1(2^{d_j},k)$ & $\bullet$   $\langle u^i+2u^th(x)\rangle$ &  $2^{2d_j(k-i)}$ \\
     &                            & \ \ ($h(x)\in ({\cal F}_j[u]/\langle u^{i-t}\rangle)^{\times}$, $t\geq 2i-k$, & \\
     &                            & \ \  $ 0\leq t<i\leq k-1)$                                 &              \\
IV.  & $\Omega_2(2^{d_j},k)$     & $\bullet$   $\langle u^i+2u^th(x) \rangle$ &  $2^{d_j(k-t)}$ \\
     &                            & \ \  ($h(x)\in ({\cal F}_j[u]/\langle u^{k-i}\rangle)^{\times}$, $t< 2i-k$, & \\
     &                            & \ \  $ 0\leq t<i\leq k-1)$        \\
V.   &  $\frac{1}{2}k(k-1)$    & $\bullet$    $\langle u^i,2u^s\rangle$ \ $(0\leq s<i\leq k-1)$ &  $2^{d_j(2k-(i+s))}$ \\
VI.   &  $(2^{d_j}-1)$    &  $\bullet$   $\langle u^i+2u^th(x), 2u^s\rangle$ &  $2^{d_j(2k-(i+s))}$ \\
     & $\cdot \Gamma(2^{d_j},k)$ & \ \  $(h(x)\in ({\cal F}_j[u]/\langle u^{s-t}\rangle)^{\times}$, & \\
     &                          & \ \  $i+s\leq k+t-1$, & \\
     &                         & \ \ $0\leq t<s<i\leq k-1)$ & \\ \hline
\end{tabular}
\end{center}}

\noindent
\textit{where}

\vskip 2mm\noindent
  $\diamond$ $\Omega_1(2^{d_j},k)=\left\{\begin{array}{ll}\frac{2^{d_j(\frac{k}{2}+1)}+
2^{d_j\cdot\frac{k}{2}}-2}{2^{d_j}-1}-(k+1), & {\rm if} \ k \ {\rm is} \ {\rm even};\cr
\frac{2(2^{d_j\cdot\frac{k+1}{2}}-1)}{2^{d_j}-1}-(k+1), & {\rm if} \ k \ {\rm is} \ {\rm odd}.\end{array}\right.$

\vskip 2mm\noindent
  $\diamond$ $\Omega_2(2^{d_j},k)=\left\{\begin{array}{ll}(2^{d_j}-1)\sum_{i=\frac{k}{2}+1}^{k-1}(2i-k)2^{d_j(k-i-1)}, & {\rm if} \ k \ {\rm is} \ {\rm even};\cr
(2^{d_j}-1)\sum_{i=\frac{k+1}{2}}^{k-1}(2i-k)2^{d_j(k-i-1)}, & {\rm if} \ k \ {\rm is} \ {\rm odd}.\end{array}\right.$

\vskip 2mm\noindent
  $\diamond$ \textit{$\Gamma(2^{d_j},k)$ can be calculated by the following recurrence formula}:

\vskip 2mm\par
   \textit{$\Gamma(2^{d_j},\rho)=0$ for $\rho=1,2,3$, $\Gamma(2^{d_j},\rho)=1$ for $\rho=4$};

\vskip 2mm\par
  \textit{$\Gamma(2^{d_j},\rho)=\Gamma(2^{d_j},\rho-1)+\sum_{s=1}^{\lfloor\frac{\rho}{2}\rfloor-1}(\rho-2s-1)2^{d_j(s-1)}$ for $\rho\geq 5$}.

\vskip 2mm \noindent
  \textit{Therefore, the number $N_{(2,d_j,k)}$ of all
distinct ideals of the ring  ${\cal K}_j[u]/\langle u^k\rangle$ is equal to}
$$N_{(2,d_j,k)}=\left\{\begin{array}{ll}\sum_{i=0}^\rho(1+4i)2^{(\rho-i)d_j} & {\rm if} \ k=2\rho; \cr & \cr
                                        \sum_{i=0}^\rho(3+4i)2^{(\rho-i)d_j} & {\rm if} \ k=2\rho+1. \end{array}\right.$$

\noindent
   \textit{Proof}
 In order to simplify the notations, we denote $\mathcal{Q}_j={\cal K}_j[u]/\langle u^k\rangle$
and ${\cal R}_j={\cal F}_j[u]/\langle u^k\rangle$. By Lemma 3.5(i), ${\cal R}_j$ is a finite chain ring with maximal
ideal $u{\cal R}_j$, the nilpotency index of $u$ is equal to $k$ and the residue field of ${\cal R}_j$ is
${\cal R}_j/(u{\cal R}_j)\cong{\cal F}_j$. Then by finite chain ring theory (cf. [9]) and $|{\cal F}_j|=2^{d_j}$, we deduce that all ideals of ${\cal R}_j$ are given by
$$\{0\}=u^k{\cal R}_j\subset u^{k-1}{\cal R}_j\subset\ldots\subset u{\cal R}_j\subset u^0{\cal R}_j={\cal R}_j$$
and $|u^l{\cal R}_j|=|{\cal F}_j|^{k-l}=2^{d_j(k-l)}$. Moreover, we have $2\mathcal{Q}_j=2{\cal R}_j$ by Lemma 3.6
where we regard ${\cal R}_j$ as a subset of $\mathcal{Q}_j$.

\par
   Let $C$ be any ideal of $\mathcal{Q}_j$. We define
\begin{center}
${\rm Tor}_0(C)=\tau(C)$, ${\rm Tor}_1(C)=\tau(\{\alpha\in \mathcal{Q}_j\mid 2\alpha\in C\})$.
\end{center}
Since $\tau$ is a surjective ring homomorphism from $\mathcal{Q}_j$ onto ${\cal R}_j$, we see that both ${\rm Tor}_0(C)$ and ${\rm Tor}_1(C)$ are ideals of ${\cal R}_j$
satisfying ${\rm Tor}_0(C)\subseteq {\rm Tor}_1(C)$. Hence there is a unique pair $(i,s)$ of integers, $0\leq s\leq i\leq k$,
such that
$${\rm Tor}_0(C)=u^i{\cal R}_j \ {\rm and} \ {\rm Tor}_1(C)=u^s{\cal R}_j.$$

\par
  Let $\tau|_C$ be the restriction of $\tau$ to $C$. Then $\tau|_C$
is a surjective ring homomorphism from $C$ onto ${\rm Tor}_0(C)$ with kernel ${\rm Ker}(\tau|_C)
=2{\rm Tor}_1(C)$,
which implies $|{\rm Ker}(\tau|_C)|=|{\rm Tor}_1(C)|$. Hence we deduce
that
\begin{equation}
|C|=|{\rm Tor}_0(C)||{\rm Tor}_1(C)|=|u^i{\cal R}_j||u^s{\cal R}_j|=2^{d_j(2k-(i+s))}
\end{equation}
by the ring isomorphism theorems and Lemma 3.5(iii). Then we have the following cases.

\par
  Case (i) $s=i$. In this case, we have $|C|=2^{2d_j(k-i)}$ by Equation (8).

\par
  (i-1) When $s=k$, then $i=k$ and $C=\{0\}=u^k\mathcal{Q}_j=\langle u^k\rangle$.

\par
  (i-2) Let $0\leq i\leq k-1$.
  By $u^i\in u^i{\cal R}_j={\rm Tor}_0(C)$, there exists $\alpha\in \mathcal{Q}_j$ such that
$u^i+2\alpha\in C$. Then by Lemma 3.6 we can assume that $\alpha\in {\cal R}_j$.

\par
  It is obvious that $\langle u^i+2\alpha\rangle=\mathcal{Q}_j(u^i+2\alpha)\subseteq C$.
Conversely, let $\xi\in C$. Since ${\rm Tor}_0(C)=u^i{\cal R}_j$, by Equation (6) and Lemma 3.6 there
exist $\beta,\gamma \in {\cal R}_j$  such that
$\xi=u^i\beta+2\gamma$, which implies
$2(\gamma-\alpha\beta)=\xi-(u^i+2\alpha)\beta\in C$ and so $\tau(\gamma-\alpha\beta)\in {\rm Tor}_1(C)=u^i{\cal R}_j$.
Hence $\tau(\gamma-\alpha\beta)=u^i\delta$ for some $\delta\in {\cal R}_j$.
From this and by ${\rm Ker}(\tau)=2{\cal Q}_j$, we deduce that
$\gamma-\alpha\beta=u^i\delta+2w$ for some $w\in {\cal Q}_j$, which implies
$2\gamma=2(\alpha\beta+u^i\delta+2w)=2\alpha\beta+2u^i\delta$.
 Therefore,
$\xi=(u^i+2\alpha)\beta+(u^i+2\alpha)\cdot 2\delta=(u^i+2\alpha)(\beta+2\delta)\in
\langle u^i+2\alpha\rangle.$
 Hence
 $$C=\langle u^i+2\alpha\rangle.$$

\par
  (i-2-1) When $i=0$, $1+2\alpha\in C$. By $(1+2\alpha)(1+2\alpha)=1$ in
$\mathcal{Q}_j$, we see that $1+2\alpha$ is an invertible element of $\mathcal{Q}_j$, which implies
$C=\mathcal{Q}_j=\langle u^0\rangle$.

\par
  (i-2-2) Let $1\leq i\leq k-1$.

\par
  Since ${\rm Tor}_1(C)=u^i{\cal R}_j$, we have $2u^i=2(u^i+2\alpha)\in C$. As $\alpha \in {\cal R}_j$, by Lemma 3.5(v) we can write $\alpha$ as $\alpha=u^i\beta$ or $\alpha=u^th+u^i\beta$ for some $\beta\in {\cal R}_j$, $0\leq t\leq i-1$
and $h\in {\cal R}_j^{\times}$.

\par
   $\diamondsuit$ When $\alpha=u^i\beta$, we have
$u^i=u^i+2\alpha-(2u^i)\beta \in C$, and so $C=\langle u^i\rangle$.

\par
  $\diamondsuit$ Let $\alpha=u^th+u^i\beta$. Then
$u^i+2u^th=u^i+2\alpha-(2u^i)\beta \in C,$
and so
$$C=\langle u^i+2u^th\rangle.$$
In this case, by $u^k=0$ we have $2u^{k-i+t}=
u^{k-i}h^{-1}(u^i+2u^th)\in C$ where $h^{-1}$ is the inverse of $h$
in ${\cal R}_j$. Therefore, $u^{k-i+t}\in {\rm Tor}_1(C)=u^i{\cal R}_j$, which implies
$k-i+t\geq i$, and hence $t\geq 2i-k$.

\par
   Now, let $h_1,h_2\in {\cal R}_j^{\times}$ and $0\leq t_1,t_2\leq k-1$ satisfying
$C=\langle u^i+2u^{t_1}h_1\rangle=\langle u^i+2u^{t_2}h_2\rangle$.
Then $2(u^{t_1}h_1-u^{t_2}h_2)=(u^i+2u^{t_1}h_1)-(u^i+2u^{t_2}h_2)\in C$,
which implies $u^{t_1}h_1-u^{t_2}h_2\in {\rm Tor}_1(C)=u^i{\cal R}_j$. From this we deduce
that $t_1=t_2=t$ and $u^t(h_1-h_2)\in u^i{\cal R}_j$. Then by finite chain ring theory (cf. [9])
and $u^t(h_1-h_2)\in u^i{\cal R}_j$, it follows that $h_1\equiv h_2$ (mod $u^{i-t}$), i.e.,
$h_1=h_2$ as elements of $({\cal R}_j/(u^{i-t}{\cal R}_j)^{\times}$. Finally, by ${\cal R}_j={\cal F}_j[u]/\langle u^k\rangle$
we have $({\cal R}_j/(u^{i-t}{\cal R}_j)^{\times}={\cal F}_j[u]/\langle u^{i-t}\rangle$ up to a natural ring isomorphism.

\par
  As stated above, we conclude that all distinct ideals $C$ of $\mathcal{Q}_j$
satisfying ${\rm Tor}_0(C)={\rm Tor}_1(C)=u^i{\cal R}_j$  are given by (I) and (III) in the table.

\vskip 2mm\par
  Case (ii) $i=k$ and $0\leq s\leq k-1$.

\par
  In this case, we have ${\rm Tor}_0(C)=\{0\}$
and $|C|=2^{d_j(2k-(k+s))}=2^{d_j(k-s)}$ by Equation (8). Moreover, by ${\rm Tor}_1(C)=u^s {\cal R}_j$ it can be easily verified that all distinct ideals of $\mathcal{Q}_j$ in this case are given by: (II) $C=\langle 2u^s\rangle$ where $0\leq s\leq k-1$.

\vskip 2mm\par
  Case (iii) $s=0$ and $1\leq i\leq k-1$.

\par
  In this case, we have $|C|=2^{d_j(2k-i)}$ by Equation (8).
Moreover, by $1\in {\rm Tor}_1(C)=u^0 {\cal R}_j$ we conclude that $2=2u^0\in C$. Then by ${\rm Tor}_0(C)=u^i {\cal R}_j$ it follows that
$C=\langle u^i,2\rangle$ immediately.

\vskip 2mm\par
   Case (iv) $1\leq s<i\leq k-1$. In this case, $|C|=2^{d_j(2k-(i+s))}$ by (8).

\par
   By ${\rm Tor}_0(C)=u^i {\cal R}_j$ and
${\rm Tor}_1(C)=u^s {\cal R}_j$ we have $2u^s\in C$ and there exists $\alpha\in {\cal R}_j$ such
that $u^i+2\alpha \in C$, which implies $\langle u^i+2\alpha, 2u^s\rangle\subseteq C$.
Conversely, let $\xi\in C$. By ${\rm Tor}_0(C)=u^i {\cal R}_j$ there exist $\beta,\gamma\in {\cal R}_j$ such that $\xi=u^i\beta+2\gamma$.
Then from $2(\gamma-\alpha\beta)=\xi-(u^i+2\alpha)\beta\in C$, we deduce
$\tau(\gamma-\alpha\beta)\in {\rm Tor}_1(C)=u^s {\cal R}_j$, which implies $\gamma-\alpha\beta=u^s \delta+2w$ for some $\delta,w\in {\cal R}_j$.
By $2\gamma=2\alpha\beta+2u^s \delta$, it follows that
$\xi=(u^i+2\alpha)\beta+(2u^s)\delta\in \langle u^i+2\alpha, 2u^s\rangle$.
Hence
$$C=\langle u^i+2\alpha, 2u^s\rangle.$$
Furthermore,  by $2u^s\in C$ and an argument similar to the proof of Case (i)
we can assume that $\alpha=0$ or $\alpha=u^th$, where $h\in {\cal R}_j^{\times}$ and $0\leq t\leq s-1$.

\par
  (iv-1) When $\alpha=0$, $C=\langle u^i, 2u^s\rangle$ which is given by (V) in the table.

\par
  (iv-2) Let $C=\langle u^i+2u^th, 2u^s\rangle$ where $h\in {\cal R}_j^{\times}$ and $0\leq t\leq s-1$.

\par
  Assume $C=\langle u^i+2u^{t_1}h_1, 2u^s\rangle=\langle u^i+2u^{t_2}h_2, 2u^s\rangle$
where $0\leq t_1,t_2\leq s-1$ and $h_1,h_2\in {\cal R}^{\times}$. Then
$2(u^{t_1}h_1-u^{t_2}h_2)=(u^i+2u^{t_1}h_1)-(u^i+2u^{t_2}h_2)\in C$, which
implies $u^{t_1}h_1-u^{t_2}h_2\in {\rm Tor}_1(C)=u^s {\cal R}_j$. From this and by an argument similar to the proof
of Case (i), we deduce that $t_1=t_2=t$ and $h_1\equiv h_2$ (mod $u^{s-t} {\cal R}_j$).
By the latter condition, we have $h=h_1=h_2\in ({\cal R}_j/(u^{s-t}{\cal R}_j))^{\times}$.
From this and by ${\cal R}_j/(u^{s-t}{\cal R}_j)={\cal F}_j[u]/\langle u^{s-t}\rangle$, we deduce
that all distinct ideals of $\mathcal{Q}_j$ in this case are given by:
$$C=\langle u^i+2u^th, 2u^s\rangle \ {\rm where} \ h\in ({\cal F}_j[u]/\langle u^{s-t}\rangle)^{\times}.$$

\par
  As $2u^{k-i+t}=u^{k-i}h^{-1}(u^i+2u^th)\in C$, we have
$u^{k-i+t}\in {\rm Tor}_1(C)=u^s {\cal R}_j$, which implies  $k-i+t\geq s$. So we have one of the following
two cases:

\par
  ($\diamondsuit$-1) $k-i+t=s$, i.e., $s-t=k-i$.

\par
  In this case, by $2u^s=u^{k-i}h^{-1}(u^i+2u^th)$ we have
$C=\langle u^i+2u^th\rangle$ and $2i>i+s=k+t$, i.e., $t<2i-k$.

\par
  Furthermore, we have
$h\in ({\cal F}_j[u]/\langle u^{k-i}\rangle)^{\times}$ and $|C|=2^{d_j(2k-(i+s))}
=2^{d_j(2k-(k+t))}=2^{d_j(k-t)}$. Hence $C$ is given by (IV) in the table.

\par
  ($\diamondsuit$-2) $k-i+t>s$, i.e., $i+s\leq k+t-1$. In this case, $C$ is given by (VI) in the table.

\par
  As stated above, we conclude that all distinct ideals and the number of elements in each ideal of $\mathcal{Q}_j$ are given by
(I)--(VI) of the table.

\par
   It is obvious that the number of ideals in (I), (II) and (V) is equal to $k+1$, $k$ and $\frac{1}{2}k(k-1)$ respectively. Then we count the number of ideals in (III), (IV) and (V) respectively.
In order to simplify the notation, we denote $q=2^{d_j}$ in the following.

\par
   First, we count the number of ideals in (III).
Let $0\leq t<i\leq k-1$ and $t\geq 2i-k$.
By Lemma 3.5(iv), it follows that $|(\mathcal{F}_j[u]/\langle u^{i-t}\rangle)^{\times}|=(2^{d_j}-1)(2^{d_j})^{(i-t)-1}=(q-1)q^{i-t-1}$.
 Then we have one of the following two cases:

\par
  ($\diamondsuit$-1) $k$ is even. In this case, $t\geq 2i-k$ if and only if $i$ and $t$ satisfy one of the following two conditions:

\par
  ($\diamondsuit$-1-1) $i\leq \frac{k}{2}$, i.e., $2i\leq k$, and $0\leq t\leq i-1$. In this case,
the number of ideals is equal to
$$N_1=\sum_{i=1}^{\frac{k}{2}}\sum_{t=0}^{i-1}|(\mathcal{F}_j[u]/\langle u^{i-t}\rangle)^{\times}|=\sum_{i=1}^{\frac{k}{2}}\sum_{t=0}^{i-1}(q-1)q^{i-t-1}=\frac{q^{\frac{k}{2}+1}-1}{q-1}-(\frac{k}{2}+1).$$

\par
  ($\diamondsuit$-1-2) $i\geq \frac{k}{2}+1$, i.e., $2i> k$, and $2i-k\leq t\leq i-1$.
In this case,
the number of ideals is equal to
$$N_2=\sum_{i=1}^{\frac{k}{2}}\sum_{t=0}^{i-1}|(\mathcal{F}_j[u]/\langle u^{i-t}\rangle)^{\times}|=\sum_{i=\frac{k}{2}+1}^{k-1}\sum_{t=2i-k}^{i-1}(q-1)q^{i-t-1}
=\frac{q^{\frac{k}{2}}-1}{q-1}-\frac{k}{2}.$$

\par
  Therefore, the number of ideals in (III) is equal to
  $\Omega_1(q,k)=N_1+N_2=\frac{q^{\frac{k}{2}+1}+
q^{\frac{k}{2}}-2}{q-1}-(k+1)$, where $k$ is even.

 ($\diamondsuit$-2) $k$ is odd. In this case, $t\geq 2i-k$ if and only if $i$ and $t$ satisfy one of the following two conditions:

\par
  ($\diamondsuit$-2-1) $i\leq \frac{k-1}{2}$, i.e., $2i\leq k$, and $0\leq t\leq i-1$. Then
the number of ideals is equal to
$N_1=\sum_{i=1}^{\frac{k-1}{2}}\sum_{t=0}^{i-1}(q-1)q^{i-t-1}=\frac{q^{\frac{k+1}{2}}-1}{q-1}-\frac{k+1}{2}$.

\par
  ($\diamondsuit$-2-2) $i\geq \frac{k-1}{2}+1$, i.e., $2i> k$, and $2i-k\leq t\leq i-1$.
Then
the number of ideals is equal to
$N_2=\sum_{i=\frac{k-1}{2}+1}^{k-1}\sum_{t=2i-k}^{i-1}(q-1)q^{i-t-1}
=\frac{q^{\frac{k+1}{2}}-1}{q-1}-\frac{k+1}{2}$.

\par
  Therefore, the number of ideals in (III) is equal to
  $\Omega_1(q,k)=N_1+N_2=\frac{2q^{\frac{k+1}{2}}-2}{q-1}-(k+1)$, where $k$ is odd.

\par
   Next we count the number of ideals in (IV).
Let $0\leq t<i\leq k-1$ and $t<2i-k$.
By Lemma 3.5(iv), it follows that $|(\mathcal{F}_j[u]/\langle u^{k-i}\rangle)^{\times}|=(q-1)q^{k-i-1}$.
Then we have one of the following two cases:

\par
  ($\diamondsuit$-1) $k$ is even. Then $t<2i-k$ if and only if $2i>k$, i.e., $i\geq \frac{k}{2}+1$,
and $0\leq t\leq 2i-k-1$. Hence the number of ideals in (IV) is equal to
\begin{eqnarray*}
\Omega_2(q,k)&=&\sum_{i=\frac{k}{2}+1}^{k-1}\sum_{t=0}^{2i-k-1}|(\mathcal{F}_j[u]/\langle u^{k-i}\rangle)^{\times}|=\sum_{i=\frac{k}{2}+1}^{k-1}\sum_{t=0}^{2i-k-1}
(q-1)q^{k-i-1}\\
&=&(q-1)\sum_{i=\frac{k}{2}+1}^{k-1}(2i-k)q^{k-i-1}.
\end{eqnarray*}

\par
  ($\diamondsuit$-2) $k$ is odd. Then $t<2i-k$ if and only if $2i>k$, i.e., $i\geq \frac{k+1}{2}$,
and $0\leq t\leq 2i-k-1$. Hence the number of ideals in (IV) is equal to
  $$\Omega_2(q,k)=\sum_{i=\frac{k+1}{2}}^{k-1}\sum_{t=0}^{2i-k-1}
(q-1)q^{k-i-1}
=(q-1)\sum_{i=\frac{k+1}{2}}^{k-1}(2i-k)q^{k-i-1}.$$

\par
   Finally, we count the number of ideals in (VI). By Lemma 3.5(iv), it follows that $|(\mathcal{F}_j[u]/\langle u^{s-t}\rangle)^{\times}|=(q-1)q^{s-t-1}$.
Hence the number of ideals in (VI) is equal to
$$\sum_{t=0}^{k-4}\sum_{t+1\leq s\leq \frac{k+t}{2}-1}\sum_{s+1\leq i\leq k+t-1-s}|(\mathcal{F}_j[u]/\langle u^{s-t}\rangle)^{\times}|
=(q-1)\Gamma(q,k),$$
where $\Gamma(q,k)=\sum_{t=0}^{k-4}\sum_{t+1\leq s\leq \frac{k+t}{2}-1}\sum_{s+1\leq i\leq k+t-1-s}q^{s-t-1}$. When $k\geq 5$,
\begin{eqnarray*}
\Gamma(q,k)&=&\sum_{t=1}^{k-4}\sum_{t+1\leq s\leq \frac{k+t}{2}-1}\sum_{s+1\leq i\leq k+t-1-s}q^{s-t-1}
   +\sum_{1\leq s\leq \frac{k}{2}-1}\sum_{s+1\leq i\leq k-1-s}q^{s-1}\\
   &=&\sum_{t^{\prime}=0}^{(k-1)-4}\sum_{t^{\prime}+1\leq s^{\prime}\leq \frac{(k-1)+t^{\prime}}{2}-1}\sum_{s^{\prime}+1\leq i^{\prime}\leq (k-1)+t^{\prime}-1-s^{\prime}}q^{s^{\prime}-t^{\prime}-1} \ \ \ \ \ \ \  \\
   &&+\sum_{s=1}^{\lfloor \frac{k}{2}\rfloor-1}(k-2s-1)q^{s-1}\\
   &=&\Gamma(q,k-1)+\sum_{s=1}^{\lfloor \frac{k}{2}\rfloor-1}(k-2s-1)q^{s-1}.
\end{eqnarray*}

\par
   If $1\leq k\leq 3$, there is no triple $(t,s,i)$ of integers satisfying $0\leq t<s<i\leq k-1$ and $i+s\leq k+t-1$.
In this case, the number of ideals in (VI) is equal to $0$. Then we set $\Gamma(q,k)=0$ for $k=1,2,3$.

\par
   If $k=4$, there is a
unique triple $(t,s,i)=(0,1,2)$ of integers satisfying $0\leq t<s<i\leq k-1$ and $i+s\leq k+t-1$. In this case, all distinct ideals in (VI) are
given by $\langle u^2+2h,2u\rangle$, where $h\in ({\cal F}_j[u]/\langle u\rangle)^{\times}={\cal F}_j^{\times}$ and
$|{\cal F}_j^{\times}|=q-1$. Then
we set $\Gamma(q,4)=1$.

\par
  Therefore, the number $N_{(2,d_j,k)}$ of ideals of $\mathcal{Q}_j$ is equal to
\begin{eqnarray*}
N_{(2,d_j,k)}&=&k+1+k+\Omega_1(q,k)+\Omega_2(q,k)+\frac{1}{2}k(k-1)+(q-1)\Gamma(q,k)\\
 &=&1+\frac{1}{2}k(k+3)+\Omega_1(q,k)+\Omega_2(q,k)+(q_j-1)\Gamma(q,k).
\end{eqnarray*}
Then by a direct calculation, we deduce that $N_{(2,d_j,2\rho)}=\sum_{i=0}^\rho(1+4i)2^{(\rho-i)d_j}$
and $N_{(2,d_j,2\rho+1)}=\sum_{i=0}^\rho(3+4i)2^{(\rho-i)d_j}$ for any positive integer $\rho$.
\hfill $\Box$

\vskip 3mm \par
   As applications of Theorem 3.7, we consider the special for $k=2,3,4,5$.

\par
   $\diamondsuit$ The number of ideals of the ring $\mathcal{K}_j[u]/\langle u^2\rangle=\mathcal{K}_j+u\mathcal{K}_j$ ($u^2=0$)
is equal to
$N_{(2,d_j,2)}=2^{d_j}+5.$
Precisely, all these ideals are given by the following table:
\begin{center}
\begin{tabular}{llll}\hline
case &  number of ideals  &  $C_j$ (ideal of ${\cal K}_j[u]/\langle u^2\rangle$)    &   $|C_j|$ \\ \hline
I.   & $3$  & $\bullet$ $\langle u^i\rangle$ \ $(i=0,1,2)$ & $2^{2d_j(2-i)}$ \\
II.   & $2$     & $\bullet$  $\langle 2u^s\rangle$ \ $(s=0,1)$ &  $2^{d_j(2-s)}$  \\
III.   & $2^{d_j}-1$ & $\bullet$   $\langle u+2h(x)\rangle$ $(h(x)\in {\cal F}_j^{\times})$ &  $2^{2d_j}$ \\
V.   &  $1$    & $\bullet$    $\langle u,2\rangle$ &  $2^{3d_j}$ \\
\hline
\end{tabular}
\end{center}

\noindent
  For example, all $7$ distinct ideals of $\mathbb{Z}_4+u\mathbb{Z}_4$ ($u^2=0$) are given by:
$\langle 1\rangle$, $\langle u\rangle$, $\langle 0\rangle$, $\langle 2\rangle$, $\langle 2u\rangle$,
$\langle u+2\rangle$ and $\langle u,2\rangle$.

\par
   $\diamondsuit$ The number of ideals of the ring $\mathcal{K}_j[u]/\langle u^3\rangle=\mathcal{K}_j+u\mathcal{K}_j+u^2\mathcal{K}_j$ ($u^3=0$)
is equal to
$N_{(2,d_j,3)}=3\cdot 2^{d_j}+7.$
Precisely, all these ideals are given by the following table:
\begin{center}
\begin{tabular}{llll}\hline
case &  number of ideals  &  $C_j$ (ideal of ${\cal K}_j[u]/\langle u^3\rangle$)    &   $|C_j|$ \\ \hline
I.   & $4$  & $\bullet$ $\langle u^i\rangle$ \ $(0\leq i\leq 4)$ & $2^{2d_j(3-i)}$ \\
II.   & $3$     & $\bullet$  $\langle 2u^s\rangle$ \ $(s=0,1,2)$ &  $2^{d_j(3-s)}$  \\
III.   & $2\cdot 2^{d_j}-2$ & $\bullet$ $\langle u+2h(x)\rangle$ $(h(x)\in {\cal F}_j^{\times})$ &  $2^{4d_j}$ \\
       &                    & $\bullet$ $\langle u^2+2uh(x)\rangle$ $(h(x)\in {\cal F}_j^{\times})$ &  $2^{2d_j}$ \\
IV.  & $2^{d_j}-1$          & $\bullet$   $\langle u^2+2h(x) \rangle$ $(h(x)\in {\cal F}_j^{\times})$ &  $2^{3d_j}$ \\
V.   &  $3$    & $\bullet$    $\langle u,2\rangle$ &  $2^{5d_j}$ \\
     &         & $\bullet$    $\langle u^2,2\rangle$ &  $2^{4d_j}$ \\
     &         & $\bullet$    $\langle u^2,2u\rangle$ &  $2^{3d_j}$ \\
\hline
\end{tabular}
\end{center}

\noindent
  For example, all $13$ distinct ideals of $\mathbb{Z}_4+u\mathbb{Z}_4+u^2\mathbb{Z}_4$ ($u^3=0$) are given by:
$\langle 1\rangle$, $\langle u\rangle$, $\langle u^2\rangle$, $\langle 0\rangle$, $\langle 2\rangle$, $\langle 2u\rangle$,
$\langle 2u^2\rangle$,
$\langle u+2\rangle$, $\langle u^2+2u\rangle$, $\langle u^2+2\rangle$,
$\langle u,2\rangle$, $\langle u^2,2\rangle$ and $\langle u^2,2u\rangle$.

\par
   $\diamondsuit$ The number of ideals of the ring $\mathcal{K}_j[u]/\langle u^4\rangle=\mathcal{K}_j+u\mathcal{K}_j+u^2\mathcal{K}_j
+u^3\mathcal{K}_j$ ($u^4=0$)
is equal to $N_{(2,d_j,4)}=2^{2d_j}+5\cdot 2^{d_j}+9$.
For example, there are $23$ distinct ideals of $\mathbb{Z}_4+u\mathbb{Z}_4+u^2\mathbb{Z}_4+u^3\mathbb{Z}_4$ ($u^4=0$).

\par
   $\diamondsuit$ The number of ideals of the ring $\mathcal{K}_j[u]/\langle u^5\rangle=\mathcal{K}_j+u\mathcal{K}_j+u^2\mathcal{K}_j
+u^3\mathcal{K}_j+u^4\mathcal{K}_j$ ($u^5=0$) is equal to
$N_{(2,d_j,5)}=3\cdot 2^{2d_j}+7\cdot 2^{d_j}+11$.
For example, there are $37$ distinct ideals of $\mathbb{Z}_4+u\mathbb{Z}_4+u^2\mathbb{Z}_4+u^3\mathbb{Z}_4+u^4\mathbb{Z}_4$ ($u^5=0$).

\vskip 3mm\par
  Finally, by Theorems 3.4 and 3.7 we deduce the following conclusion.

\vskip 3mm \noindent
   {\bf Corollary 3.8} \textit{Every cyclic code $\mathcal{C}$
over $R=\mathbb{Z}_4[u]/\langle u^k\rangle$ of length $n$ can be constructed by the following two steps}:

\vskip 2mm\par
   {\bf Step 1} \textit{For each integer $j$, $1\leq j\leq r$, choose an ideal
$C_j$ of the ring $\mathcal{K}_j[u]/\langle u^k\rangle$ listed by Theorem 3.7};

\vskip 2mm\par
   {\bf Step 2} \textit{Set $\mathcal{C}=\bigoplus_{j=1}^r\Psi(e_j(x)C_j)$}.

\vskip 2mm\par
 \textit{Moreover, the number of all cyclic codes over $R$
of length $n$ is equal to $\prod_{j=1}^rN_{(2,d_j,k)}$}.

\section{Dual codes and self-duality of cyclic codes over $R$ of length $n$}
\noindent
   In this section, we give the dual code of each cyclic code over the ring $R=\mathbb{Z}_{4}[u]/\langle u^k\rangle$
of odd length $n$, and then investigate the self-duality of these codes.

\par
   For any $\alpha=(\alpha_0,\alpha_1,\ldots,\alpha_{n-1}), \beta=(\beta_0,\beta_1,\ldots,\beta_{n-1})\in R^{n}$,
where $\alpha_j,\beta_j\in R$ and $0\leq j\leq n-1$, recall that
the usual \textit{Euclidian inner product} of $\alpha$ and $\beta$ are defined by
$[\alpha,\beta]_E=\sum_{j=0}^{n-1}\alpha_j\beta_j\in R$.
It is known that $[-,-]_E$ is a symmetric and non-degenerate bilinear form on the $R$-module
$R^{n}$, i.e.,
$$[\alpha,\beta]_E=0, \ \forall \beta\in R^{n} \Longrightarrow \beta=0$$
for any $\alpha\in R^{n}$. Let $C$ be a linear code over $R$ of length ${n}$, i.e.,
an $R$-submodule of $R^{n}$. The \textit{Euclidian dual code}
of $C$ is defined by $C^{\bot_E}=\{\alpha\in R^{n}\mid [\alpha,\beta]_E=0, \ \forall
\beta\in C\}$, and $C$ is said to be \textit{self-dual} if $C=C^{\bot_E}$. As usual, we will identify $\alpha\in R^{n}$
with $\alpha(x)=\sum_{j=0}^{n-1}\alpha_jx^j\in R[x]/\langle x^{n}-1\rangle$ in this paper.
Then we define
$$\varrho(\alpha(x))=\alpha(x^{-1})=\alpha_0+\sum_{j=1}^{n-1}\alpha_jx^{n-j}, \ \forall \alpha(x)\in R[x]/\langle x^{n}-1\rangle.$$
It is clear that $\varrho$ is a ring automorphism of $R[x]/\langle x^{n}-1\rangle$ satisfying
$\varrho^{-1}=\varrho$. Now, by a direct calculation we get the following lemma.

\vskip 3mm \noindent
  {\bf Lemma 4.1} \textit{Let $\alpha,\beta\in R^{n}$.
Then $[\alpha,\beta]_E=0$  if $\alpha(x)\varrho(\beta(x))=0$
in the ring $R[x]/\langle x^{n}-1\rangle$}.

\vskip 3mm \par
   By Lemma 3.1, the map $\Psi$ defined in Section 3 is a ring isomorphism
from ${\cal A}[u]/\langle u^k\rangle$ onto $R[x]/\langle x^{n}-1\rangle$. Now, let
$\sigma=\Psi^{-1}\varrho \Psi$. Then $\sigma$ is a ring automorphism of
${\cal A}[u]/\langle u^k\rangle$ such that
the following diagram commutates:
$$\begin{array}{ccc} \mathcal{A}[u]/\langle u^k\rangle & \stackrel{\sigma}{\longrightarrow} &  \mathcal{A}[u]/\langle u^k\rangle \cr
  \Psi \downarrow &  & \ \ \ \downarrow \Psi \cr
R[x]/\langle x^n-1\rangle & \stackrel{\varrho}{\longrightarrow} &  R[x]/\langle x^n-1\rangle,
\end{array}$$
It is obvious that $\sigma^{-1}=\sigma$ and the restriction of $\sigma$
to ${\cal A}$ is a ring automorphism of ${\cal A}$. We still denote this automorphism by $\sigma$. Then
\begin{equation}
\sigma(a(x))=a(x^{-1})=a_0+\sum_{i=1}^{n-1}a_ix^{n-i}, \ \forall a(x)=\sum_{i=0}^{n-1}a_ix^{i}\in {\cal A}.
\end{equation}

\par
   Let $1\leq j\leq r$. By Equations (5) and the definition of $e_j(x)$ in Section 3, we have
\begin{equation}
\sigma(e_j(x))=v_j(x^{-1})F_j(x^{-1})=1-w_j(x^{-1})f_j(x^{-1}) \ {\rm in} \ {\cal A}.
\end{equation}
\par
   For any polynomial $f(x)=\sum_{l=0}^da_lx^l\in \mathbb{Z}_{4}[x]$ of degree $d\geq 1$, recall that
the \textit{reciprocal polynomial} of $f(x)$ is defined as $\widetilde{f}(x)=x^df(\frac{1}{x})=\sum_{l=0}^da_lx^{d-l}$, and
 $f(x)$ is said to be \textit{self-reciprocal} if $\widetilde{f}(x)=\delta f(x)$ for some $\delta \in \mathbb{Z}_{4}^{\times}=\{1,3\}$. By
$x^n-1=f_1(x)f_2(x)\ldots f_r(x)$ given in Section 3, we have
$$x^{n}-1=3\widetilde{f}_1(x)\widetilde{f}_2(x)\ldots \widetilde{f}_r(x).$$
Since $f_1(x),f_2(x),\ldots,f_r(x)$ are pairwise coprime monic basic irreducible polynomials in $\mathbb{Z}_{4}[x]$,
 $\widetilde{f}_1(x),\widetilde{f}_2(x),\ldots, \widetilde{f}_r(x)$  are pairwise coprime basic irreducible polynomials in $\mathbb{Z}_{4}[x]$ as well. Hence for each integer $j$, $1\leq j\leq r$,
there is a unique integer $j^{\prime}$, $1\leq j^{\prime}\leq r$, such that
$$\widetilde{f}_j(x)=\delta_jf_{j^{\prime}}(x) \ {\rm for} \ {\rm some} \ \delta_j\in \mathbb{Z}_{4}^{\times}.$$

\par
  We assume that $f_j(x)=\sum_{l=0}^{d_j}c_lx^l$ where $c_j\in \mathbb{Z}_{4}$. Then
$x^{d_j}f_j(x^{-1})=\widetilde{f}_j(x)$. From this, by Equation (10) and $x^{n}=1$ in ${\cal A}$ we deduce
\begin{eqnarray*}
\sigma(e_j(x))&=&1-x^{n-({\rm deg}(v_j(x))+d_j)}(x^{{\rm deg}(v_j(x))}v_j(x^{-1}))
(x^{d_j}f_j(x^{-1}))\\
  &=&1-x^{n-({\rm deg}(v_j(x))+d_j)}\widetilde{v}_j(x)\widetilde{f}_j(x)\\
  &=&1-b_j(x)f_{j^{\prime}}(x)
\end{eqnarray*}
where $b_j(x)=\delta_jx^{n-({\rm deg}(v_j(x))+d_j)}\widetilde{v}_j(x)\in {\cal A}$.
Similarly, by Equation (5) in Section 3 it
follows that $\sigma(e_j(x))=g_j(x)F_{j^{\prime}}(x)$ for some $g_j(x)\in {\cal A}$. Then from these,
by Equations (10) and the definition of $e_{j^{\prime}}(x)$ we deduce that
$$\sigma(e_j(x))=e_{j^{\prime}}(x).$$

\par
   As stated above, we see that
for each integer $j$, $1\leq j\leq r$, there is a unique integer $j^{\prime}$, $1\leq j^{\prime}\leq r$, such that $\sigma(e_j(x))=
e_{j^{\prime}}(x)$. We still use $\sigma$ to denote this map $j\mapsto j^{\prime}$; i.e., $\sigma(e_j(x))=e_{\sigma(j)}(x)$.
Whether $\sigma$ denotes the ring automorphism of ${\cal A}$ or this map on the set $\{1,\ldots,r\}$ is determined by context.
The next lemma shows the compatibility of the two uses of $\sigma$.

\vskip 3mm \noindent
  {\bf Lemma 4.2} \textit{With the notations above, we have the following}:

\vskip 2mm \par
   (i) \textit{$\sigma$ is a permutation on $\{1,\ldots,r\}$ satisfying $\sigma^{-1}=\sigma$}.

\vskip 2mm \par
   (ii) \textit{After a rearrangement of $e_1(x),\ldots,e_r(x)$ there are integers $\lambda,\epsilon$ such that
$\sigma(j)=j$ for all $j=1,\ldots,\lambda$ and $\sigma(\lambda+l)=\lambda+\epsilon+l$ for all $l=1,\ldots,\epsilon$, where $\lambda\geq 1,
\epsilon\geq 0$ and $\lambda+2\epsilon=r$}.

\vskip 2mm\par
   (iii) \textit{For each integer $j$, $1\leq j\leq r$, there is a unique element $\delta_j$ of
$\mathbb{Z}_{4}^{\times}$ such that $\widetilde{f}_j(x)=\delta_j f_{\sigma(j)}(x)$}.

\vskip 2mm \par
   (iv) \textit{For any integer $j$, $1\leq j\leq r$, $\sigma(e_j(x))=e_{\sigma(j)}(x)$ in the ring ${\cal A}$,
and $\sigma({\cal A}_{j})={\cal A}_{\sigma(j)}$. Then $\sigma$ induces a ring isomorphism from ${\cal A}_{j}[u]/\langle u^k\rangle$
 onto ${\cal A}_{\sigma(j)}[u]/\langle u^k\rangle$, which is still denoted by $\sigma$, in the natural way}:
$$\sum_{l=0}^{k-1}a_l(x)u^i\mapsto \sum_{l=0}^{k-1}\sigma(a_l(x))u^i=\sum_{l=0}^{k-1}a_l(x^{-1})u^i,
\ \forall a_l(x)\in {\cal A}_{j}, \ 0\leq l \leq k-1.$$

\vskip 3mm \noindent
  \textit{Proof.} (i)--(iii) follow from the definition of the map $\sigma$, and
(iv) follows from that ${\cal A}_j=e_j(x){\cal A}$ immediately.
\hfill $\Box$

\vskip 3mm \par
  Since ${\cal A}=\bigoplus_{j=1}^r{\cal A}_j$ by Lemma 3.2(ii), it follows that
${\cal A}[u]=\bigoplus_{j=1}^r{\cal A}_j[u]$, and hence ${\cal A}[u]/\langle u^k\rangle=\bigoplus_{j=1}^r{\cal A}_j[u]/\langle u^k\rangle$.

\vskip 3mm \noindent
  {\bf Lemma 4.3} \textit{Let $\alpha(x)=\sum_{j=1}^r\alpha_j(x), \beta(x)=\sum_{j=1}^r\beta_j(x)\in {\cal A}[u]/\langle u^k\rangle$,
where $\alpha_j(x), \beta_j(x)\in{\cal A}_j[u]/\langle u^k\rangle$. Then}
$\alpha(x)\sigma(\beta(x))=\sum_{j=1}^r\alpha_j(x)\sigma(\beta_{\sigma(j)}(x)).$

\vskip 3mm \noindent
  \textit{Proof.} By Lemma 4.2(iv) and $\sigma^{-1}=\sigma$, it follows that
$$\sigma(\beta_{\sigma(j)}(x))\in \sigma ({\cal A}_{\sigma(j)}[u]/\langle u^k\rangle)={\cal A}_{\sigma^2(j)}[u]/\langle u^k\rangle={\cal A}_j[u]/\langle u^k\rangle.$$
Hence $\alpha_j(x)\sigma(\beta_{\sigma(j)}(x))\in {\cal A}_j[u]/\langle u^k\rangle$ for all $j$. If
$l\neq \sigma(j)$, then $j\neq\sigma(l)$ and so ${\cal A}_j{\cal A}_{\sigma(l)}=\{0\}$ by Lemma 3.2(ii),
which implies
$$\alpha_j(x)\sigma(\beta_l(x))\in ({\cal A}_j[u]/\langle u^k\rangle)\cdot
({\cal A}_{\sigma(l)}[u]/\langle u^k\rangle)
=\{0\}$$
Hence

\par
\ \ \  $\alpha(x)\sigma(\beta(x))=\sum_{j=1}^r\sum_{l=1}^r\alpha_j(x)\sigma(\beta_{l}(x))=\sum_{j=1}^r\alpha_j(x)\sigma(\beta_{\sigma(j)}(x))$.
\hfill $\Box$

\vskip 3mm \par
   By Lemma 3.2(iii), we know that $\varphi_j: b(x)\mapsto e_j(x)b(x)$ mod $x^n-1$
($\forall b(x)\in \mathcal{K}_j$) is a ring isomorphism from the Galois ring $\mathcal{K}_j=\mathbb{Z}_4[x]/\langle f_j(x)\rangle$
onto $\mathcal{A}_j$.

\vskip 3mm \noindent
  {\bf Lemma 4.4} \textit{Using the notations above, for any $1\leq j\leq r$ we have}

\vskip 2mm \par
  (i) \textit{Let $\sigma_j=\varphi_{\sigma(j)}^{-1}\sigma|_{\mathcal{A}_j}\varphi_j$ where $\sigma|_{\mathcal{A}_j}$
is the restriction of $\sigma$ to $\mathcal{A}_j$.
Then
$\sigma_j$ is a ring isomorphism from ${\cal K}_j$ onto ${\cal K}_{\sigma(j)}$ such that the following diagram commutes}
$$\begin{array}{ccc} \ \ \ \ {\cal K}_j=\mathbb{Z}_{4}[x]/\langle f_j(x)\rangle & \stackrel{\sigma_j}{\longrightarrow} &  {\cal K}_{\sigma(j)}=\mathbb{Z}_{4}[x]/\langle f_{\sigma(j)}(x)\rangle \cr
  \varphi_j  \downarrow &  & \ \ \ \downarrow \varphi_{\sigma(j)} \cr
 \ \ \ \ {\cal A}_j=e_j(x){\cal A} & \stackrel{\sigma|_{\mathcal{A}_j}}{\longrightarrow} &  {\cal A}_{\sigma(j)}=e_{\sigma(j)}(x){\cal A}
\end{array}.$$
\textit{Specifically, for any $b(x)\in {\cal K}_j$ we have $\sigma_j(b(x))=b(x^{-1})\in {\cal K}_{\sigma(j)}$,
where}
$$b(x^{-1})\equiv b(x^{n-1}) \ ({\rm mod} \ f_{\sigma(j)}(x)).$$

\vskip 2mm \par
  (ii) \textit{$\sigma_j$ induces a ring isomorphism from ${\cal K}_j[u]/\langle u^{k}\rangle$ onto ${\cal K}_{\sigma(j)}[u]/\langle u^{k}\rangle$,
which will be still denoted by $\sigma_j$,
defined in the natural way that}:
$$\xi\mapsto \sum_{l=0}^{k-1}\sigma_j(\alpha_l)u^l,
\ {\rm i.e.,} \ \sigma_j(\xi)=\sum_{l=0}^{k-1}\sigma_j(\alpha_l)u^l,$$
\textit{for any $\xi=\sum_{l=0}^{k-1}\alpha_lu^l$ $\in {\cal K}_j[u]/\langle u^k\rangle$
where $\alpha_0,\alpha_1,\ldots,\alpha_{k-1}\in {\cal K}_j$}.

\vskip 3mm \noindent
  \textit{Proof} (i) It follows from Lemma 3.2(iii) and Lemma 4.2 (iv).

\par
  (ii) It follows from (i) immediately.
\hfill $\Box$

\vskip 3mm \par
   Now, we give the dual code of each cyclic code over $R$ of length $n$.

\vskip 3mm \noindent
   {\bf Theorem 4.5} \textit{Let ${\cal C}$ be a cyclic code
over $R$ of length $n$ with canonical form decomposition
${\cal C}=\bigoplus_{j=1}^r\Psi(e_j(x)C_j)$, where $C_j$ is an ideal of the ring ${\cal K}_j[u]/\langle u^k\rangle$ given by Theorem 3.7. Then
the dual code ${\cal C}^{\bot_E}$ of ${\cal C}$ is given by}
$${\cal C}^{\bot_E}=\bigoplus_{j=1}^r\Psi(e_j(x)D_j),$$
\textit{where $D_j$ is an ideal of ${\cal K}_j[u]/\langle u^k\rangle$ given by one of the following eight cases for all
$j=1,\ldots,r$}:
\begin{center}
\begin{tabular}{lll}\hline
case &  $C_j$  &  $D_{\sigma(j)}$  \\ \hline
1.   & $\langle u^i\rangle$ \ $(0\leq i\leq k)$ &  $\langle u^{k-i}\rangle$  \\
2.   &  $\langle 2u^s\rangle$ \ ($0\leq s\leq k-1$) & $\langle u^{k-s},2\rangle$ \\
3.   & $\langle u^i+2u^th(x)\rangle$ & $\langle u^{k-i}+2u^{k+t-2i}h(x^{-1})\rangle$ \\
     & ($h(x)\in ({\cal F}_j[u]/\langle u^{i-t}\rangle)^{\times}$, $t\geq 2i-k$, & \\
     & $0\leq t<i\leq k-1$) & \\
4.   & $\langle u^i+2h(x)\rangle$  & $\langle u^i+2h(x^{-1})\rangle$ \\
     & $(h(x)\in ({\cal F}_j[u]/\langle u^{k-i}\rangle)^{\times}$, $2i>k$, & \\
     & $0<i\leq k-1$) & \\
5.   & $\langle u^i+2u^{t}h(x)\rangle$ & $\langle u^{i-t}+2h(x^{-1}),2u^{k-i}\rangle$  \\
     & $(h(x)\in ({\cal F}_j[u]/\langle u^{k-i}\rangle)^{\times}$, $t<2i-k$,   &  \\
     & $1\leq t<i\leq k-1$)     & \\
6.   & $\langle u^{i},2u^{s}\rangle$ \ $(0\leq s<i\leq k-1)$ & $\langle u^{k-s},2u^{k-i}\rangle$ \\
7.   & $\langle u^{i}+2h(x), 2u^{s}\rangle$ & $\langle u^{k-s}+2u^{k-i-s}h(x^{-1})\rangle$ \\
     & $(h(x)\in ({\cal F}_j[u]/\langle u^s\rangle)^{\times}$, $i+s\leq k-1$,  &  \\
     & $1\leq s<i\leq k-1$) & \\
8. & $\langle u^{i}+2u^{t}h(x), 2u^{s}\rangle$ & $\langle u^{k-s}+2u^{k+t-i-s}h(x^{-1}),$ \\
   & $(h(x)\in ({\cal F}_j[u]/\langle u^{s-t}\rangle)^{\times}$,  &  $2u^{k-i}\rangle$ \\
   &  $i+s\leq k+t-1$,  & \\
   & $1\leq t<s<i\leq k-1)$ & \\ \hline
\end{tabular}
\end{center}

\noindent
\textit{in which $h(x^{-1})\in {\cal F}_{\sigma(j)}[u]/\langle u^k\rangle$ satisfying $h(x^{-1})\equiv
h(x^{n-1}) \ ({\rm mod} \ \overline{f}_{\sigma(j)}(x))$}.

\vskip 3mm \noindent
  \textit{Proof} Let $1\leq j\leq r$, and
$E_j$ be an ideal of ${\cal K}_j[u]/\langle u^k\rangle$ given by one of the following eight cases:

\par
  (i) $E_j=\langle u^{k-i}\rangle$,  if $C_j=\langle u^i\rangle$, where $0\leq i\leq k$.

\par
  (ii) $E_j=\langle u^{k-s},2\rangle$,  if $C_j=\langle 2u^s\rangle$, where $0\leq s\leq k-1$.

\par
  (iii)  $E_j=\langle u^{k-i}+2u^{k+t-2i}h(x)\rangle$, if $C_j=\langle u^i+2u^th(x)\rangle$ where $h(x)\in ({\cal F}_j[u]/\langle u^{i-t}\rangle)^{\times}$,
 $0\leq t<i\leq k-1$ and $t\geq 2i-k$.

\par
  (iv) $E_j=\langle u^i+2h(x)\rangle$,  if $C_j=\langle u^i+2h(x)\rangle$ where $h(x)\in ({\cal F}_j[u]/\langle u^{k-i}\rangle)^{\times}$,
$0<i\leq k-1$ and $2i>k$.

\par
  (v) $E_j=\langle u^{i-t}+2h(x),2u^{k-i}\rangle$,  if $C_i=\langle u^i+2u^{t}h(x)\rangle$ where
  $h(x)\in ({\cal F}_j[u]/\langle u^{k-i}\rangle)^{\times}$,
$1\leq t<i\leq k-1$ and $t<2i-k$.

\par
  (vi) $E_j=\langle u^{k-s},2u^{k-i}\rangle$, if $C_j=\langle u^{i},2u^{s}\rangle$, where
$0\leq s<i\leq k-1$.

\par
 (vii) $E_j=\langle u^{k-s}+2u^{k-i-s}h(x)\rangle$,  if $C_j=\langle u^{i}+2h(x), 2u^{s}\rangle$, where $h(x)\in ({\cal F}_j[u]/\langle u^s\rangle)^{\times}$,
$1\leq s<i\leq k-1$, $i+s\leq k-1$.

\par
  (viii) $E_j=\langle u^{k-s}+2u^{k+t-i-s}h(x), 2u^{k-i}\rangle$, when $C_j=\langle u^{i}+2u^{t}h(x), 2u^{s}\rangle$, where $h(x)\in ({\cal F}_j[u]/\langle u^{s-t}\rangle)^{\times}$,
$1\leq t<s<i\leq k-1$, $i+s\leq k+t-1$.

\vskip 2mm \noindent
  Then by a direct calculation we deduce that
$C_jE_j=\{a(x)b(x)\mid a(x)\in C_j, \ b(x)\in E_j\}=\{0\}$.
Further, we have $|C_j||E_j|=2^{2d_jk}$ by Theorem 3.7.

\par
   Since $\sigma_j$ is a ring isomorphism from ${\cal K}_j[u]/\langle u^k\rangle$ onto ${\cal K}_{\sigma(j)}[u]/\langle u^k\rangle$ define
by Lemma 4.4(ii), $\sigma_j(E_j)$ is an ideal of ${\cal K}_{\sigma(j)}[u]/\langle u^k\rangle$
and $|\sigma_j(E_j)|=|E_j|$. We
denote $D_{\sigma(j)}=\sigma_j(E_j)$. Let $l=\sigma(j)$. Then $j=\sigma(l)$ as $\sigma^{-1}=\sigma$.

\par
   Now, set
${\cal D}=\sum_{j=1}^r\Psi(e_{\sigma(j)}(x) D_{\sigma(j)})
=\bigoplus_{l=1}^r\Psi(e_l(x)D_{l})$, where $D_{l}=\sigma_{\sigma(l)}(E_{\sigma(l)})$ is an
ideal of ${\cal K}_l[u]/\langle u^k\rangle$ since $\sigma_{\sigma(l)}=\sigma_l^{-1}: \mathcal{K}_{\sigma(l)}[u]/\langle u^k\rangle
\rightarrow \mathcal{K}_{l}[u]/\langle u^k\rangle$.
By Theorem 3.4, we see that ${\cal D}$ is a cyclic code over $R$ of length $n$.

\par
   For any $1\leq j\leq r$, by Lemma 3.2(iii) we deduce that $e_{\sigma(j)}(x) D_{\sigma(j)}$ is an ideal of
${\cal A}_{\sigma(j)}[u]/\langle u^k\rangle$. Moreover, we have $\sigma(D_{\sigma(j)})=\sigma^2(E_j)=E_j$,
and $\sigma(e_{\sigma(j)}(x))=e_j(x)$ by Lemma 4.2(iv). From these
and by Lemma 4.3, we deduce that
\begin{eqnarray*}
{\cal C}\cdot \sigma({\cal D})&=&\left(\sum_{j=1}^r\Psi(e_j(x)C_j)\right)\cdot \sigma\left(\sum_{l=1}^r\Psi(e_l(x)D_l)\right)\\
    &=&\Psi\left(\sum_{j=1}^r(e_j(x)C_j)\cdot \sigma\left(e_{\sigma(j)}(x)D_{\sigma(j)}\right)\right)\\
    &=&\Psi\left(\sum_{j=1}^r(e_j(x)C_j) (e_j(x)E_j)\right)\\
    &=&\Psi\left(\sum_{j=1}^re_j(x)(C_j E_j)\right)\\
    &=&\{0\},
\end{eqnarray*}
which implies ${\cal D}\subseteq {\cal C}^{\bot_E}$ by Lemma 4.1.
  On the other hand, by Theorems 3.4 and 3.7, we have
\begin{eqnarray*}
|{\cal C}||{\cal D}|&=&(\prod_{j=1}^r|C_j|)(\prod_{j=1}^r|E_j|)=\prod_{j=1}^r(|C_j||E_j|)
=\prod_{j=1}^r2^{2d_jk}\\
 &=&2^{2k\sum_{j=1}^rd_j}=(4^{k})^{n}=|R^n|.
\end{eqnarray*}
Since
$R=\mathbb{Z}_{4}[u]/\langle u^k\rangle$ is a finite Frobenius ring, by the theory of linear codes over Frobenius rings (See Dougherty et al. [7])
we deduce that ${\cal C}^{\bot_E}={\cal D}$.

\par
  Finally, we give the generator set of $D_{\sigma(j)}=\sigma_j(E_j)$, $1\leq j\leq r$.
By Lemma 4.4, we have
$\sigma_j(u^l)=u^l$ for all $0\leq l\leq k$, and for any $h(x)\in {\cal F}_j[u]/\langle u^k\rangle$
$$\sigma_j(h(x))=h(x^{-1})\equiv
h(x^{n-1}) \ ({\rm mod} \ \overline{f}_{\sigma(j)}(x)),$$
which implies
$$\sigma_j(2u^lh(x))=2u^lh(x^{-1}) \ {\rm with} \ h(x^{-1})\in {\cal F}_{\sigma(j)}[u]/\langle u^k\rangle.$$
Then the conclusions follows from (i)--(viii) immediately.
\hfill $\Box$

\vskip 3mm \par
   Finally, using Theorems 3.4, 3.7 and 4.5 we list all distinct self-dual cyclic
codes over the ring $R$ of length $n$ by the following Theorem.

\vskip 3mm
\noindent
  {\bf Theorem 4.6} \textit{Using the notations in Lemma 4.2$({\rm ii})$, all
distinct self-dual cyclic
codes over the ring $R$ of length $n$ are give by}:
$${\cal C}=\left(\bigoplus_{j=1}^\lambda \Psi(e_j(x)C_j)\right)\oplus
\left(\bigoplus_{j=\lambda+1}^{\lambda+\epsilon}\Psi(e_{j}(x)C_{j})\oplus\Psi(e_{j+\epsilon}(x)C_{j+\epsilon})\right),$$
\textit{where $C_j$ is an ideal of ${\cal K}_j[u]/\langle u^k\rangle$ determined by the following conditions}:

\vskip 2mm\par
  (A) \textit{If $1\leq j\leq \lambda$, $C_j$ is determined by the following conditions}:

\vskip 2mm\par
  (A-i) \textit{When $k$ is even, $C_j$ is given by one of the following six cases}:

  \vskip 2mm \par
  (A-i-1) \textit{$C_j=\langle u^{\frac{k}{2}}\rangle$}.

\vskip 2mm \par
  (A-i-2) \textit{$C_j=\langle 2\rangle$}.

\vskip 2mm \par
  (A-i-3) \textit{$C_j=\langle u^{\frac{k}{2}}+2u^{t}h(x)\rangle$, where
$h(x)\in ({\cal F}_j[u]/\langle u^{\frac{k}{2}-t}\rangle)^{\times}$ satisfying
$h(x)+h(x^{-1})\equiv 0$ $($mod $\overline{f}_j(x)$$)$, and $0\leq t\leq \frac{k}{2}-1$}.

\vskip 2mm \par
  (A-i-4) \textit{$C_j=\langle u^{i}+2h(x)\rangle$, where
$h(x)\in ({\cal F}_j[u]/\langle u^{k-i}\rangle)^{\times}$ satisfying
$h(x)+h(x^{-1})\equiv 0$ $($mod $\overline{f}_j(x)$$)$, and $\frac{k}{2}+1\leq i\leq k-1$}.

\vskip 2mm \par
  (A-i-5) \textit{$C_j=\langle u^i,2u^{k-i}\rangle$, where
$\frac{k}{2}+1<i\leq k-1$}.

\vskip 2mm \par
  (A-i-6) \textit{$C_j=\langle u^{i}+2u^{t}h(x), 2u^{k-i}\rangle$, where
$1\leq t<k-i$, $\frac{k}{2}+1\leq i\leq k-1$ and
$h(x)\in ({\cal F}_j[u]/\langle u^{k-i-t}\rangle)^{\times}$ satisfying
$h(x)+h(x^{-1})\equiv 0$ $($mod $\overline{f}_j(x)$$)$}.

\vskip 2mm\par
  (A-ii) \textit{When $k$ is odd, $C_j$ is given by one of the following four cases}:

\vskip 2mm \par
  (A-ii-1) \textit{$C_j=\langle 2\rangle$}.

\vskip 2mm \par
  (A-ii-2) \textit{$C_j=\langle u^{i}+2h(x)\rangle$, where
$h(x)\in ({\cal F}_j[u]/\langle u^{k-i}\rangle)^{\times}$ satisfying
$h(x)+h(x^{-1})\equiv 0$ $($mod $\overline{f}_j(x)$$)$, and $\frac{k+1}{2}\leq i\leq k-1$}.

\vskip 2mm \par
  (A-ii-3) \textit{$C_j=\langle u^i,2u^{k-i}\rangle$, where
$\frac{k+1}{2}<i\leq k-1$}.

\vskip 2mm \par
  (A-ii-4) \textit{$C_j=\langle u^{i}+2u^{t}h(x), 2u^{k-i}\rangle$, where
$1\leq t<k-i$, $\frac{k+1}{2}\leq i\leq k-1$ and
$h(x)\in ({\cal F}_j[u]/\langle u^{k-i-t}\rangle)^{\times}$ satisfying
$h(x)+h(x^{-1})\equiv 0$ $($mod $\overline{f}_j(x)$$)$}.

\vskip 2mm \par
  (B) \textit{If $j=\lambda+l$ where $1\leq l\leq \epsilon$, then $(C_j, C_{j+\epsilon})$ is given by
one of the following eight cases}:
{\small\begin{center}
\begin{tabular}{lll}\hline
case &  $C_j$  &  $C_{j+\epsilon}$  \\ \hline
B-1.   & $\langle u^i\rangle$ \ $(0\leq i\leq k)$ &  $\langle u^{k-i}\rangle$  \\
B-2.   &  $\langle 2u^s\rangle$ \ ($0\leq s\leq k-1$) & $\langle u^{k-s},2\rangle$ \\
B-3.   & $\langle u^i+2u^th(x)\rangle$ & $\langle u^{k-i}+2u^{k+t-2i}h(x^{-1})\rangle$ \\
     & ($h(x)\in ({\cal F}_j[u]/\langle u^{i-t}\rangle)^{\times}$, & \\
     &  $t\geq 2i-k, 0\leq t<i\leq k-1$) & \\
B-4.   & $\langle u^i+2h(x)\rangle$  & $\langle u^i+2h(x^{-1})\rangle$ \\
     & $(h(x)\in ({\cal F}_j[u]/\langle u^{k-i}\rangle)^{\times}$, & \\
     & $2i>k, 0<i\leq k-1$) & \\
B-5.   & $\langle u^i+2u^{t}h(x)\rangle$ & $\langle u^{i-t}+2h(x^{-1}),2u^{k-i}\rangle$  \\
     & $(h(x)\in ({\cal F}_j[u]/\langle u^{k-i}\rangle)^{\times}$,    &  \\
     & $t<2i-k, 1\leq t<i\leq k-1$)     & \\
B-6.   & $\langle u^{i},2u^{s}\rangle$ & $\langle u^{k-s},2u^{k-i}\rangle$ \\
       & $(0\leq s<i\leq k-1)$  & \\
B-7.   & $\langle u^{i}+2h(x), 2u^{s}\rangle$ & $\langle u^{k-s}+2u^{k-i-s}h(x^{-1})\rangle$ \\
     & $(h(x)\in ({\cal F}_j[u]/\langle u^s\rangle)^{\times}$,   &  \\
     & $i+s\leq k-1, 1\leq s<i\leq k-1$) & \\
B-8. & $\langle u^{i}+2u^{t}h(x), 2u^{s}\rangle$ & $\langle u^{k-s}+2u^{k+t-i-s}h(x^{-1}),2u^{k-i}\rangle$ \\
   & $(h(x)\in ({\cal F}_j[u]/\langle u^{s-t}\rangle)^{\times}$,  &   \\
   &  $i+s\leq k+t-1$,  & \\
   & $1\leq t<s<i\leq k-1)$ & \\ \hline
\end{tabular}
\end{center}}

\noindent
\textit{in which $h(x^{-1})\in {\cal F}_{j+\epsilon}[u]/\langle u^k\rangle$ satisfying $h(x^{-1})\equiv
h(x^{n-1}) \ ({\rm mod} \ \overline{f}_{j+\epsilon}(x))$}.


\section{Quasi-cyclic codes of index $4$ over $\mathbb{Z}_4$ derived from cyclic codes over $\mathbb{Z}_4[u]/\langle u^4\rangle$}
\noindent
In this section, let $R=\mathbb{Z}_4[u]/\langle u^4\rangle$. We consider quasi-cyclic codes of length $4n$ and index $4$ derived from cyclic codes over $R$
of length $n$ where $n$ is odd.

Using the notations of Section 3, let $\mathcal{A}=\mathbb{Z}_4[x]/\langle x^n-1\rangle$.
It is well-known that ideals of the ring $\mathcal{A}$ are identified with cyclic codes of
length $n$. Obviously,
$$\mathcal{A}^4=\{(\xi_0,\xi_1,\xi_2,\xi_3)\mid \xi_0,\xi_1,\xi_2,\xi_3\in \mathcal{A}\}$$
is an $\mathcal{A}$-module with componentwise addition and scalar multiplication from elements of
$\mathcal{A}$. Then $\mathcal{A}$-submodules of $\mathcal{A}^4$ are identified as
quasi-cyclic codes of length $4n$ and index $4$ (cf. Cui [6] or Cao [5], for example).

\par
   Since $\mathcal{A}$ is a subring of $\mathcal{A}[u]/\langle u^4\rangle=\mathcal{A}+u\mathcal{A}+u^2\mathcal{A}+u^3\mathcal{A}$
($u^4=0$), $\mathcal{A}[u]/\langle u^4\rangle$ is an $\mathcal{A}$-module. Now, we define a map
$\Upsilon: \mathcal{A}[u]/\langle u^4\rangle\rightarrow \mathcal{A}^4$ by
$$\Upsilon:\xi_0+u\xi_1+u^2\xi_2+u^3\xi_3\mapsto (\xi_3,\xi_2+\xi_3,\xi_1+\xi_2+\xi_3,\xi_0+\xi_1+\xi_2+\xi_3)$$
($\forall \xi_0,\xi_1,\xi_2,\xi_3\in \mathcal{A}$). Then it can be easily verify
that $\Upsilon$ is an $\mathcal{A}$-module isomorphism from $\mathcal{A}[u]/\langle u^4\rangle$
onto $\mathcal{A}^4$.

\par
  Let $\Psi^{-1}$ be the inverse of the ring isomorphism $\Psi$ (see Lemma 3.1) from $\mathcal{A}[u]/\langle u^4\rangle$ onto
$R[x]/\langle x^n-1\rangle$. Then $\Psi^{-1}$ is a ring isomorphism from $R[x]/\langle x^n-1\rangle$
onto $\mathcal{A}[u]/\langle u^4\rangle$. Since $\mathbb{Z}_4$ is a subring of both $R$ and $\mathcal{A}$,
by Lemma 3.1 we deduce that $\Upsilon\Psi^{-1}$ is a $\mathbb{Z}_4$-isomorphism from
$R[x]/\langle x^n-1\rangle$ onto $\mathcal{A}^4$.

\par
  On the other hand, it is clear that $R[x]/\langle x^n-1\rangle$ is a free $R$-module
of rank $n$ with a basis $\{1,x,\ldots, x^{n-1}\}$. As $R$ is a finite commutative local ring,
any $R$-submodule of $R[x]/\langle x^n-1\rangle$ must be finitely generated. In the following,
we identify $R[x]/\langle x^n-1\rangle$ with $R^n$ under the natural $R$-module isomorphism:
$$\alpha_0+\alpha_1x+\ldots+\alpha_{n-1}x^{n-1}\mapsto (\alpha_0,\alpha_1,\ldots,\alpha_{n-1})$$
$(\forall \alpha_0,\alpha_1,\ldots,\alpha_{n-1}\in R)$.

\par
  Recall that $R$-submodules of $R^n$ are in fact linear codes over $R$ of length $n$.
Let $S$ be a linear code over $R$ of length $n$. A matrix $G$ is called a \textit{generator matrix} for $S$ if every codeword in $S$
is an $R$-linear combination of the row vectors of $G$ and
any row vector of $S$ can not be written as an $R$-linear combination of the other row vectors of $G$ (cf. [9] Definition 3.1).
We assume that $G$ is a matrix over $R$ of size $m\times n$, i.e,. $G\in {\rm M}_{m\times n}(R)$. Since each element $\alpha$ of $R$ can be uniquely expressed as:
$\alpha=a_0+ua_1+u^2a_2+u^3a_3$ where $a_0,a_1,a_2,a_3\in \mathbb{Z}_4$, the matrix $G$ can be uniquely write as:
$$G=G_0+uG_1+u^2G_2+u^3G_3, \ G_0,G_1,G_2,G_3\in {\rm M}_{m\times n}(\mathbb{Z}_4).$$
Especially, every vector $\upsilon$ of $R^m$ can be uniquely write as:
$$\upsilon=\alpha+u\beta+u^2\gamma+u^3\delta, \ \alpha,\beta,\gamma,\delta\in \mathbb{Z}_4^m.$$
   Then we have the following theorem.

\vskip 3mm \noindent
  {\bf Theorem 5.1} \textit{Let $\mathcal{C}$ be a cyclic code over $R$ of length $n$,
i.e., $\mathcal{C}$ is  an ideal of the ring $R[x]/\langle x^n-1\rangle$. Denote}
$$D=\Upsilon\Psi^{-1}(\mathcal{C})=\Upsilon(\Psi^{-1}(\mathcal{C})).$$

\par
  (i) \textit{$D$ is a quasi-cyclic code over $\mathbb{Z}_4$ of length $4n$ and index $4$}.

\par
  (ii) \textit{If $G_{\mathcal{C}}=G_0+uG_1+u^2G_2+u^3G_3$ is a generator matrix of $\mathcal{C}$ as an
$R$-submodule of $R^n$, where $G_0,G_1,G_2,G_3\in {\rm M}_{m\times n}(\mathbb{Z}_4)$ for some positive integer $m$. Then as a $\mathbb{Z}_4$-submodule
of $\mathbb{Z}_4^{4n}$, a generator matrix of $D$ is given by}
$$G_{D}=\left(\begin{array}{rrrr}G_3 & G_2+G_3 & G_1+G_2+G_3 & G_0+G_1+G_2+G_3\cr
  G_2 & G_1+G_2 & G_0+G_1+G_2 & G_0+G_1+G_2\cr
  G_1 & G_0+G_1 & G_0+G_1 & G_0+G_1\cr
  G_0 & G_0 & G_0 & G_0\end{array}\right).$$

\vskip 3mm \noindent
  \textit{Proof} (i) Since $\mathcal{C}$ is  an ideal of the ring $R[x]/\langle x^n-1\rangle$
and $\Psi^{-1}$ is a ring isomorphism from $R[x]/\langle x^n-1\rangle$
onto $\mathcal{A}[u]/\langle u^4\rangle$, we see that $\Psi^{-1}(\mathcal{C})$ is an ideal
of the ring $\mathcal{A}[u]/\langle u^4\rangle$, which implies that $\Psi^{-1}(\mathcal{C})$ is an
$\mathcal{A}$-submodule of $\mathcal{A}[u]/\langle u^4\rangle$. Hence $D=\Upsilon(\Psi^{-1}(\mathcal{C}))$
is a quasi-cyclic code over $\mathbb{Z}_4$ of length $4n$ and index $4$ since
$\Upsilon$ is an $\mathcal{A}$-module isomorphism from $\mathcal{A}[u]/\langle u^4\rangle$
onto $\mathcal{A}^4$.

\par
  (ii) Since $G_{\mathcal{C}}$ is a a generator matrix of $\mathcal{C}$ as an
$R$-submodule of $R^n$, it follows that
\begin{eqnarray*}
\mathcal{C}&=&\{(\alpha+u\beta+u^2\gamma+u^3\delta) G_{\mathcal{C}}\mid \alpha,\beta,\gamma,\delta\in \mathbb{Z}_4^m\}\\
  &=&\{\alpha G_0+u(\alpha G_1+\beta G_0)+u^2(\alpha G_2+\beta G_1+\gamma G_0)\\
  &&+u^3(\alpha G_3+\beta G_2+\gamma G_1+\delta G_0)\mid \alpha,\beta,\gamma,\delta\in \mathbb{Z}_4^m\}.
\end{eqnarray*}
From this and by the definition of $\Upsilon$, we deduce that
\begin{eqnarray*}
&&\Upsilon((\alpha+u\beta+u^2\gamma+u^3\delta) G_{\mathcal{C}})\\
  &=&(\alpha G_3+\beta G_2+\gamma G_1+\delta G_0,\\
  &&(\alpha G_2+\beta G_1+\gamma G_0)+(\alpha G_3+\beta G_2+\gamma G_1+\delta G_0),\\
  &&(\alpha G_1+\beta G_0)+(\alpha G_2+\beta G_1+\gamma G_0)+(\alpha G_3+\beta G_2+\gamma G_1+\delta G_0),\\
  &&\alpha G_0+(\alpha G_1+\beta G_0)+(\alpha G_2+\beta G_1+\gamma G_0)+(\alpha G_3+\beta G_2+\gamma G_1+\delta G_0)\\
  &=&\underline{w}G_D,
\end{eqnarray*}
where $\underline{w}=(\alpha,\beta,\gamma,\delta)\in \mathbb{Z}_4^{4m}$,
which implies $D=\{\underline{w}G_D\mid \underline{w}\in \mathbb{Z}_4^{4m}\}$.
Hence $G_D$ is a generator matrix of $D$ as a $\mathbb{Z}_4$-submodule
of $\mathbb{Z}_4^{4n}$.
\hfill $\Box$

\vskip 3mm \noindent
  {\bf Remark} In Theorem 5.1, the quasi-cyclic code $D$ over $\mathbb{Z}_4$ of length $4n$ and index $4$
is completely determined by the ideal $\Psi^{-1}(\mathcal{C})$ of $\mathcal{A}[u]/\langle u^4\rangle$.
All distinct ideals of $\mathcal{A}[u]/\langle u^4\rangle$ can be provided by Lemma 3.3
and Theorem 3.7.


\section{An Example}
\noindent
    We consider cyclic codes of length $7$ over
$R=\mathbb{Z}_4[u]/\langle u^k\rangle=\mathbb{Z}_4+u\mathbb{Z}_4+\ldots+u^{k-1}\mathbb{Z}_4$ ($u^k=0$)
for $k=2,3,4,5$, respectively. In this case, we have $n=7$.

\par
  It is known that $x^7-1=f_1(x)f_2(x)f_3(x)$
where
$$f_1(x)=x+3, \ f_2(x)=x^3+2x^2+x+3, \ f_3(x)=x^3+3x^2+2x+3$$
are basic irreducible monic polynomials in $\mathbb{Z}_4[x]$
satisfying $\widetilde{f}_1(x)=3f_1(x)$ and $\widetilde{f}_2(x)=3f_3(x)$.
Hence $r=3$, $d_1=1$, $d_2=d_3=3$, $\lambda=1$ and $\epsilon=1$.

\par
  $\bullet$ First, we calculate the number of cyclic codes of length $7$ over
$R=\mathbb{Z}_4[u]/\langle u^k\rangle$ for $k=2,3,4,5$.
   By Theorem 3.7 and Corollary 3.8, the number of cyclic codes over
$R$ of length $7$ is equal to $\mathcal{N}_{(7,k)}=\prod_{j=1}^3N_{(2,d_j,k)}=N_{(2,1,k)}N_{(2,3,k)}^2$,
where
$$N_{(2,1,k)}=\left\{\begin{array}{ll}7, & {\rm if} \ k=2;\cr
   13, & {\rm if} \ k=3;\cr
   23, & {\rm if} \ k=4;\cr
   37, & {\rm if} \ k=5,\end{array}\right. \ {\rm and} \
N_{(2,3,k)}=\left\{\begin{array}{ll}13, & {\rm if} \ k=2;\cr
   31, & {\rm if} \ k=3;\cr
   113, & {\rm if} \ k=4;\cr
   259, & {\rm if} \ k=5.\end{array}\right.$$
Therefore, we deduce the following:

\vskip 2mm\par
  $\diamond$ The number of cyclic codes of length $7$ over
$\mathbb{Z}_4+u\mathbb{Z}_4$ ($u^2=0$) is equal to $\mathcal{N}_{(7,2)}=7\cdot 13^2=1,183$;

\vskip 2mm\par
  $\diamond$ The number of cyclic codes of length $7$ over
$\mathbb{Z}_4+u\mathbb{Z}_4+u^2\mathbb{Z}_4$ ($u^3=0$) is equal to $\mathcal{N}_{(7,3)}=13\cdot 31^2=12,493$;

\vskip 2mm\par
  $\diamond$ The number of cyclic codes of length $7$ over
$\mathbb{Z}_4+u\mathbb{Z}_4+u^2\mathbb{Z}_4+u^3\mathbb{Z}_4$ ($u^4=0$) is equal to $\mathcal{N}_{(7,4)}=23\cdot 113^2=293,687$;

\vskip 2mm\par
  $\diamond$ The number of cyclic codes of length $7$ over
$\mathbb{Z}_4+u\mathbb{Z}_4+u^2\mathbb{Z}_4+u^3\mathbb{Z}_4+u^4\mathbb{Z}_4$ ($u^5=0$) is equal to $\mathcal{N}_{(7,5)}=37\cdot 259^2=2,481,997$.

\vskip 2mm\par
  $\bullet$ Next, we list all cyclic codes of length $7$ over
$R=\mathbb{Z}_4+u\mathbb{Z}_4+u^2\mathbb{Z}_4+u^3\mathbb{Z}_4$ ($u^4=0$). For each $j=1,2,3$,
let $F_j(x)=\frac{x^7-1}{f_j(x)}\in \mathbb{Z}_4[x]$. Then we find polynomials
$v_j(x),w_j(x)\in \mathbb{Z}_4[x]$ satisfying $v_j(x)F_j(x)+w_j(x)f_j(x)=1$, and set
$e_j(x)\equiv v_j(x)F_j(x)$ (mod $x^7-1$). Precisely, we have

\vskip 2mm\par
  $e_1(x)=3\,{x}^{6}+3\,{x}^{5}+3\,{x}^{4}+3\,{x}^{3}+3\,{x}^{2}+3\,x+3$;

\vskip 2mm\par
  $e_2(x)=2\,{x}^{6}+2\,{x}^{5}+3\,{x}^{4}+2\,{x}^{3}+3\,{x}^{2}+3\,x+1$.

\vskip 2mm\par
  $e_3(x)=3\,{x}^{6}+3\,{x}^{5}+2\,{x}^{4}+3\,{x}^{3}+2\,{x}^{2}+2\,x+1$;

\vskip 2mm\par
  Using the notations of Section 3, we have

\par
  $\diamondsuit$ $\mathcal{K}_1=\mathbb{Z}_4[x]/\langle f_1(x)\rangle=\mathbb{Z}_4$,
$\mathcal{F}_1=\mathbb{F}_2[x]/\langle \overline{f}_1(x)\rangle=\mathbb{F}_2=\{0,1\}$
and $\mathcal{F}_1[u]/\langle u^l\rangle=\mathbb{F}_2+u\mathbb{F}_2+\ldots+u^{l-1}\mathbb{F}_2$ for
any $l=1,2,3$. Precisely, we have
$$\mathcal{F}_1^{\times}=\{1\} \ {\rm and} \
(\mathcal{F}_1+u\mathcal{F}_1)^{\times}=\{1,1+u\}.$$

\par
  $\diamondsuit$ For $j=2,3$, $\mathcal{K}_j=\mathbb{Z}_4[x]/\langle f_j(x)\rangle=\{a_0+a_1x+a_2x^2\mid a_0,a_1,a_2\in \mathbb{Z}_4\}$,
$\mathcal{F}_j=\mathbb{F}_2[x]/\langle \overline{f}_j(x)\rangle=\{b_0+b_1x+b_2x^2\mid b_0,b_1,b_2\in \mathbb{F}_2\}$,
where

\vskip 2mm\par
 $\diamond$  $\overline{f}_3(x)=x^3+x+1$, $\overline{f}_2(x)=x^3+x^2+1$;

\vskip 2mm\par
 $\diamond$ $\mathcal{F}_j[u]/\langle u^l\rangle=\mathcal{F}_j+u\mathcal{F}_j+\ldots+u^{l-1}\mathcal{F}_j$, $1\leq l\leq 4$.

\vskip 2mm\noindent
  Precisely, we have
$$\mathcal{F}_j^{\times}=\{b_0+b_1x+b_2x^2\mid (b_0,b_1,b_2)\neq (0,0,0), \ b_0,b_1,b_2\in \mathbb{F}_2\}$$
and
\begin{eqnarray*}
(\mathcal{F}_j+u\mathcal{F}_j)^{\times}&=&\{\alpha+u\beta\mid \alpha\in \mathcal{F}_j^{\times}, \ \beta\in \mathcal{F}_j\}\\
  &=&\{b_0+b_1x+b_2x^2+u(c_0+c_1x+c_2x^2)\\
  &&\mid (b_0,b_1,b_2)\neq (0,0,0), \ b_0,b_1,b_2,c_0,c_1,c_2\in \mathbb{F}_2\}.
\end{eqnarray*}

\par
  By by Theorem 3.7 and Corollary 3.8, all distinct cyclic codes of length $7$ over
$R$ are given by
$${\cal C}=\Psi(e_1(x)C_1)\oplus \Psi(e_2(x)C_2)\oplus \Psi(e_3(x)C_3) \ {\rm with} \ |{\cal C}|=|C_1||C_2||C_3|,$$
where

\par
  $\diamond$ $C_1$ is an ideal of the ring ${\cal K}_1[u]/\langle u^4\rangle$ listed by the following table:
{\small\begin{center}
\begin{tabular}{llll}\hline
case &  number of ideals  &  $C_1$ (ideal of $\mathbb{Z}_4[u]/\langle u^4\rangle$)    &   $|C_1|$ \\ \hline
I.   & $5$  & $\bullet$ $\langle u^i\rangle$ \ $(0\leq i\leq 4)$ & $4^{4-i}$ \\
II.  & $4$     & $\bullet$  $\langle 2u^s\rangle$ \ $(0\leq s\leq 3)$ &  $2^{4-s}$  \\
III. & $5$ & $\bullet$   $\langle u+2\rangle$ &  $4^{3}$ \\
     &     & $\bullet$   $\langle u^2+2\rangle$, $\langle u^2+2(1+u)\rangle$, $\langle u^2+2u\rangle$ &  $4^{2}$ \\
     &     & $\bullet$   $\langle u^3+2u^2\rangle$  &  $4$ \\
IV.  & $2$  & $\bullet$   $\langle u^3+2 \rangle$  &  $2^{4}$ \\
     &      & $\bullet$   $\langle u^3+2u \rangle$ &  $2^{3}$ \\
V.   &  $6$    & $\bullet$    $\langle u^i,2u^s\rangle$  $(0\leq s<i\leq 3)$ &  $2^{8-(i+s)}$ \\
VI.   &  $1$    &  $\bullet$   $\langle u^2+2, 2u\rangle$ &  $2^{5}$ \\
 \hline
\end{tabular}
\end{center}}

\par
  $\diamond$ For $j=2,3$, $C_j$ is an ideal of the ring ${\cal K}_j[u]/\langle u^4\rangle$ listed by the following table:
{\small\begin{center}
\begin{tabular}{llll}\hline
case &  number of ideals  &  $C_j$ (ideal of ${\cal K}_j[u]/\langle u^4\rangle$)    &   $|C_j|$ \\ \hline
I.   & $5$  & $\bullet$ $\langle u^i\rangle$ \ $(0\leq i\leq 4)$ & $4^{3\cdot(4-i)}$ \\
II.  & $4$     & $\bullet$  $\langle 2u^s\rangle$ \ $(0\leq s\leq 3)$ &  $2^{3\cdot(4-s)}$  \\
III. & $77$ & $\bullet$   $\langle u+2h(x)\rangle$ ($h(x)\in {\cal F}_j^{\times}$) &  $4^{9}$ \\
     &      & $\bullet$   $\langle u^2+2h(x)\rangle$ ($h(x)\in ({\cal F}_j+u{\cal F}_j)^{\times}$) &  $4^{6}$ \\
     &      & $\bullet$   $\langle u^2+2uh(x)\rangle$ ($h(x)\in {\cal F}_j^{\times}$) &  $4^{6}$ \\
     &      & $\bullet$   $\langle u^3+2u^2h(x)\rangle$ ($h(x)\in {\cal F}_j^{\times}$) &  $4^{3}$ \\
IV.  & $14$  & $\bullet$   $\langle u^3+2h(x) \rangle$ ($h(x)\in {\cal F}_j^{\times}$) &  $2^{12}$ \\
     &                     & $\bullet$   $\langle u^3+2uh(x) \rangle$ ($h(x)\in {\cal F}_j^{\times}$) &  $2^{9}$ \\
V.   &  $6$    & $\bullet$    $\langle u^i,2u^s\rangle$  $(0\leq s<i\leq 3)$ &  $2^{3\cdot (8-(i+s))}$ \\
VI.   &  $7$    &  $\bullet$   $\langle u^2+2h(x), 2u\rangle$ ($h(x)\in {\cal F}_j^{\times}$)&  $2^{15}$ \\
 \hline
\end{tabular}
\end{center}}

\vskip 2mm\par
  $\bullet$ Now, we list all distinct self-dual cyclic codes of length $7$
over $R=\mathbb{Z}_4+u\mathbb{Z}_4+u^2\mathbb{Z}_4+u^3\mathbb{Z}_4$ ($u^4=0$).
By Theorem 4.6, all distinct self-dual cyclic codes of length $7$
over $R$ are given by:
$${\cal C}=\Psi(e_1(x)C_1)\oplus \Psi(e_2(x)C_2)\oplus \Psi(e_3(x)C_3),$$
where

\par
  $\diamond$ $C_1$ is one of the following $7$ ideals:
$$\langle u^2\rangle, \langle 2\rangle, \langle u^2+2\rangle, \langle u^2+2(1+u)\rangle, \langle u^2+2u\rangle, \langle u^3+2\rangle,
\langle u^3,2u\rangle.$$

\par
  $\diamond$ The pair $(C_2,C_3)$ of ideals are given by the following table:
{\small\begin{center}
\begin{tabular}{llll}\hline
 number of pairs $(C_2,C_3)$ & $C_2$  &  $C_{3}$  \\ \hline
 $5$ & $\langle u^i\rangle$  \  ($0\leq i\leq 4$) &  $\langle u^{k-i}\rangle$  \\
 $4$  &  $\langle 2u^s\rangle$  \  ($0\leq s\leq 3$) & $\langle u^{4-s},2\rangle$ \\
 $7$  &  $\langle u+2h(x)\rangle$ \ ($h(x)\in {\cal F}_2^{\times}$) & $\langle u^3+2u^2h(x^{-1})\rangle$  \\
 $56$  &  $\langle u^2+2h(x)\rangle$ \ ($h(x)\in ({\cal F}_2+u{\cal F}_2)^{\times}$) &
          $\langle u^2+2h(x^{-1})\rangle$ \\
 $7$   &  $\langle u^2+2uh(x)\rangle$ \ ($h(x)\in {\cal F}_2^{\times}$) & $\langle u^2+2uh(x^{-1})\rangle$ \\
 $7$   &  $\langle u^3+2u^2h(x)\rangle$ \ ($h(x)\in {\cal F}_2^{\times}$) & $\langle u+2h(x^{-1})\rangle$  \\
 $7$  &  $\langle u^3+2h(x)\rangle$  ($h(x)\in {\cal F}_2^{\times}$) & $\langle u^3+2h(x^{-1})\rangle$   \\
 $7$  &  $\langle u^3+2uh(x)\rangle$ \ ($h(x)\in {\cal F}_2^{\times}$) & $\langle u^2+2h(x^{-1}),2u\rangle$   \\
 $6$  & $\langle u^{i},2u^{s}\rangle$ \ ($0\leq s<i\leq 3$) & $\langle u^{4-s},2u^{4-i}\rangle$ \\
 $7$  & $\langle u^2+2h(x), 2u\rangle$ \ ($h(x)\in {\cal F}_2^{\times}$) & $\langle u^3+2uh(x^{-1})\rangle$  \\
 \hline
\end{tabular}
\end{center}}

\noindent
in which

\par
   $\diamondsuit$  if $h(x)=b_0+b_1x+b_2x^2\in {\cal F}_2^{\times}$, where $b_0,b_1,b_2\in \mathbb{F}_2$
satisfying $(b_0,b_1,b_2)$ $\neq (0,0,0)$, then
$h(x^{-1})\equiv b_0+b_1x^6+b_2x^5 \ ({\rm mod} \ x^3+x^2+1)$, i.e.,
$$h(x^{-1})=(b_0+b_2)+(b_1+b_2)x+b_1x^2\in {\cal F}_3^{\times};$$

\par
   $\diamondsuit$  if $h(x)=b_0+b_1x+b_2x^2+u(c_0+c_1x+c_2x^2)\in ({\cal F}_2+u{\cal F}_2)^{\times}$,
where $b_0,b_1,b_2,c_0,c_1,c_2\in \mathbb{F}_2$
satisfying $(b_0,b_1,b_2)\neq (0,0,0)$, then $h(x^{-1})\in ({\cal F}_3+u{\cal F}_3)^{\times}$ is given by
$$h(x^{-1})=(b_0+b_2)+(b_1+b_2)x+b_1x^2+u\left((c_0+c_2)+(c_1+c_2)x+c_1x^2\right).$$

\par
  Therefore, the number of self-dual cyclic codes of length $7$
over $R$ is equal to $7\cdot 113=791$.

\vskip 2mm\par
  $\bullet$ Finally, we list some new optimal quasi-cyclic code with better parameter over $\mathbb{Z}_4$ of length $28$ and index $4$ obtained by Theorem 5.1 as the following table (existing optical parameter for $\mathbb{Z}_4$ code in [2]):
{\small\begin{center}
\begin{tabular}{llll}\hline
 $C_1$ & $C_2$ & $C_3$  & $\Upsilon(\Psi^{-1}(\mathcal{C}))$   \\ \hline
   $u^4$ & $u^3$ & $u^4$ &$[28,2^6,24]$\\
    $u^4$ & $u^4$ & $u^3+2(x^2+1)u^2$ & $[28,2^6,24]$\\
   $u^4$ & $u^4$ & $u^3+2xu^2$ & $[28,2^6,24]$ \\
    $u^4$ & $u^4$ & $u^3+2(x^2+x)u^2$ & $[28,2^6,24]$\\
   $u^4$ & $u^4$ & $u^3+2u^2$ & $[28,2^6,24]$ \\
    $u^4$ & $u^4$ & $u^3+2(x^2+1)u^2$  & $[28,2^6,24]$\\
    $u^4$ & $u^4$ & $u^3+2x^2u^2$ & $[28,2^6,24]$\\
$u^4$ & $u^4$ & $u^3+2(x^2+x+1)u^2$ & $[28,2^6,24]$\\
$u^3$ & $u^4$ & $u^3+2x^2u^2$ & $[28,2^8,20]$ \\
 $u^3$ & $u^4$ & $u^3+2(x^2+x)u^2$ & $[28,2^8,20]$\\

$u^3$ & $u^4$ & $u^3+2xu^2$ & $[28,2^8,20]$ \\
 $u^3$ & $u^4$ & $u^3+2(x^2+1)u^2$ & $[28,2^8,20]$\\

$u^3$ & $u^4$ & $u^3+2(x+1)u^2$ & $[28,2^8,20]$ \\
$u^3$ & $u^4$ & $u^3+2(x^2+x+1)u^2$ & $[28,2^8,20]$\\
$u^3+2u^2$ & $u^4$ & $u^3+2x^2u^2$ & $[28,2^8,20]$\\
$u^3+2u^2$ & $u^4$ & $u^3+2xu^2$ & $[28,2^8,20]$\\
$u^3+2u^2$ & $u^4$ & $u^3+2(x^2+x)u^2$ & $[28,2^8,20]$\\
$u^3+2u^2$ & $u^4$ & $u^3+2(x^2+1)u^2$ & $[28,2^8,20]$\\
$u^3+2u^2$ & $u^4$ & $u^3+2(x+1)u^2$ & $[28,2^8,20]$\\
$u^3+2u^2$ & $u^4$ & $u^3+2(x^2+x+1)u^2 $ & $[28,2^8,20]$\\
 \hline
\end{tabular}
\end{center}}

\noindent
in which
$\Upsilon(\Psi^{-1}(\mathcal{C}))$ is the quasi-cyclic code over $\mathbb{Z}_4$ of length $28$ and index $4$
derived from $\mathcal{C}=\Psi(e_1(x)C_1\oplus e_2(x)C_2 \oplus e_3(x)C_3)$ (mod $x^7-1$) using Theorem 5.1,
and $\Upsilon(\Psi^{-1}(\mathcal{C}))$ has basic parameters $[28, M, d]$ if
$|\Upsilon(\Psi^{-1}(\mathcal{C}))|=|C_1||C_2||C_3|=M$
and the minimum Lee distance of $\Upsilon(\Psi^{-1}(\mathcal{C}))$ is equal to $d$.


\section{Conclusions and further research}
\noindent
We have developed a theory for cyclic codes of length $n$ over the finite commutative locasl ring $R=\mathbb{Z}_{4}[u]/\langle u^k\rangle=\mathbb{Z}_{4}
+u\mathbb{Z}_{4}+\ldots+u^{k-1}\mathbb{Z}_{4}$ ($u^k=0$) for any integer $k\geq 2$ and odd positive integer $n$,
including the enumeration and construction of
these codes, the dual code and self-duality for each of these codes. These codes enjoy a rich algebraic structure compared
to arbitrary linear codes (which makes the search process much simpler).
Obtaining some bounds for minimal distance such as BCH-like of a cyclic code over the ring $R$ by just looking at the representation of such codes are future topics of interest.

\section*{Acknowledgements}
This research is supported in part by the Research Fund for the Doctoral Program of Higher Education of China (No. 20120161110017).



\end{document}